\documentclass[a4paper, 12pt]{article}
\usepackage[italian, german, english]{babel}
\usepackage{amsmath}
\usepackage{amsfonts}
\usepackage{amsthm}
\usepackage{mathrsfs}
\usepackage{latexsym}
\usepackage[dvips]{graphicx}
\newcommand{\field}[2]{\mathbb{#1}^{#2}}
\newcommand{\abs}[1]{\lvert#1\rvert}
\newcommand{\norm}[1]{\lVert#1\rVert}
\newcommand{\deriv}[2]{\partial_{#1}^{#2}}
\newcommand{\weakstar}{\overset{*}{\rightharpoonup}}
\newcommand{\heps}{\widehat{H}_{\varepsilon}}
\newtheorem{theo}{Theorem}
\newtheorem{prop}{Proposition}
\newtheorem{lem}{Lemma}
\newtheorem{cor}{Corollary}
\newcommand{\gammatheta}{\Gamma_{\vartheta}}

\theoremstyle{remark}
\newtheorem{rem}{Remark}

\title{Quantum graphs as holonomic constraints}
\author{Gianfausto Dell'Antonio$^{\textrm{a)}}$ and Lucattilio Tenuta$^{\textrm{b)}}$}
\date{}

\begin{document}

\maketitle

a) Dipartimento di Matematica, Universit\`a di Roma I ``La Sapienza'', \\Piazzale Aldo Moro 2, 00185, Roma, Italy 

\verb+e-mail: dellantonio@mat.uniroma1.it+\\

b) Mathematisches Institut, Universit\"at T\"ubingen, Auf der Morgenstelle 10, 72076, T\"ubingen, Germany 

\verb+e-mail: lucattilio.tenuta@uni-tuebingen.de+

\begin{abstract}
We consider the dynamics on a quantum graph as the limit of the dynamics generated by a one-particle Hamiltonian in $\field{R}{2}$ with a potential having a deep strict minimum on the graph, when the width of the well shrinks to zero. For a generic graph we prove convergence outside the vertices to the free dynamics on the edges. For a simple model of a graph with two edges and one vertex, we prove convergence of the dynamics to the one generated by the Laplacian with Dirichlet boundary conditions in the vertex. 
\end{abstract}

\textbf{I. INTRODUCTION}

Several physical systems, like electronic nanostructures or periodic solids made up of aromatic molecules, display the common feature that the motion of the electrons can be thought of as being confined in one or more directions by a strong potential barrier, which prevents the particles from escaping from the structure and allows propagation in the remaining (free) directions.

In solid state physics, a concrete example of this phenomenon is given by quantum wires and carbon nanotubes, where the high purity achieved in fabrication techniques and the weakness of electron-phonon interaction give rise to ballistic transport. Therefore, using the strong-coupling method, one can in first approximation model the interaction of electrons with the crystal assuming that they move freely with an effective mass $m$ (a more detailed discussion of the physical hypotheses on which this approximation is based can be found in Duclos and Exner (1995), Londergan \emph{et al.} (1999)).

Taking into account the strong potential barrier which keeps them confined, their dynamics is then given by the one-particle Hamiltonian
\begin{equation}\label{heps}
\heps := -\frac{1}{2}\Delta + \frac{1}{2\varepsilon^{2}}W(q),
\end{equation}
where we have chosen suitable units so that $\hbar$ and the effective mass are equal to $1$. 

The parameter $\varepsilon$ is the natural small parameter of the problem and is linked to the ratio $l/L$, where $L$ is the characteristic length of the wire along the free direction where the electrons can propagate and $l$ is the analogous length in the confined directions.

In general, the function $W$ is assumed to be zero on the quantum wire and strictly positive outside, so that when $\varepsilon$ becomes small one expects the electron to be better and better confined to the wire. For this reason, $W$ is called the \emph{constraining potential}.

Our aim in this paper is to analyze the dynamics generated by \eqref{heps} in the limit $\varepsilon \to 0$ for two-dimensional systems constrained to a singular one-dimensional manifold given by a graph $\Gamma$.

We consider an explicit form of the constraining potential, i. e., the distance of a point from the graph. This potential is obviously continuous, but does not belong to $\textrm{C}^{1}(\field{R}{2})$, because of the vertices. Outside the vertices, the potential is a quadratic function of the coordinates, so it can be regarded as the first non zero term of the Taylor expansion of more general functions which are zero on the graph, and whose gradient is zero on the edges. This is the natural class of constraining potentials considered in the literature to model holonomic constraints in classical and quantum mechanics (Bornemann (1998), Froese and Herbst (2001), and references therein). We will argue that the results we find can be generalized to a wider class of potentials, whose Taylor expansion near the graph contains higher order terms.

As it is clear from the brief remarks given above, this case is relevant for solid state physics in order to determine the leading order behavior for the dynamics of electrons moving in a \emph{branched} nanostructure, made up of several wires which meet in a crossing region represented by the vertices of the graph in the limit $\varepsilon\to 0$.

\begin{figure}
\begin{center}
\includegraphics[scale=0.2]{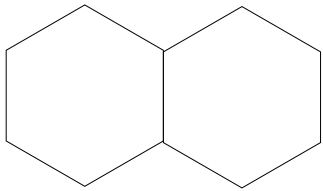}\caption{Skeleton of naphthalene molecule}\label{naphthalene}
\end{center}
\end{figure}

Another important field where equation \eqref{heps} can be applied is given by theoretical chemistry, in the so-called Quantum Network Model (QNM) (for a recent review which contains also some comparison with experimental data see Amovilli \emph{et al.} (2004) and references therein). This model is used to study the motion of valence electrons (also called $\pi$-electrons) in aromatic molecules or periodic solids like graphene. In first approximation, they are thought to move freely through the skeleton of the molecule determined by $\sigma$-electrons, which create a potential keeping $\pi$-electrons confined to the molecular structure. The most famous example, described in figure \ref{naphthalene}, is probably naphthalene molecule, first studied by Ruedenberg and Scherr (1953).

The characteristic feature shared by the examples presented is that we always expect the limiting dynamics to take place on a graph, on which we have to define a suitable Hamiltonian. The couple made up of the graph and a differential or pseudo-differential self-adjoint operator defined on it is usually called quantum graph (a detailed review on this topic is given in Kuchment (2002, 2004, 2005)). 

The main problem one runs into if one wants to describe the physical systems mentioned above by a quantum graph is that there are a host of self-adjoint Hamiltonians which can be defined on it. 

If, for example, one assumes that the dynamics outside the vertices is free, i. e., that the Hamiltonian is the Laplacian $-d^{2}/dx^{2}$, where $x$ is the natural arc length coordinate on the edges of the graph, and that the graph has $1$ vertex and $n$ edges, then the possible self-adjoint extensions are determined by $n^{2}$ parameters (Kostrykin and Schrader (1999)). 

To eliminate this ambiguity, a natural procedure is to consider the quantum graph as a limit of a more realistic model. We call the method employing a strong constraining potential a \emph{soft approximation} of the graph. As we have pointed out above, this is a physically reasonable approximation, which, for a smooth constraint, has been investigated, e. g., in Belov \emph{et al.} (2004), Dell'Antonio and Tenuta (2004), Froese and Herbst (2001). As far as authors' knowledge is concerned, the remark by Kuchment (2002) that ``the graph case (i. e., in the presence of vertices) has not been explored'' is still largely true.

Another appropriate choice, which we call \emph{rigid approximation} of quantum graphs, is to consider a ``thickened graph'', composed in the simplest cases of thin tubes of radius $\varepsilon$, which has the same topology as the original graph and reduces to it when $\varepsilon$ goes to zero. It is reasonable to suppose that the motion of the electron in this thickened structure is free, but one needs to specify boundary conditions to get a well-defined Laplacian. The most natural choice is to use Dirichlet boundary conditions, which correspond to an infinite constraining potential barrier. Some light on this case has been shed only very recently by Post (2005), who considered thickened graphs which are strictly \emph{smaller} than the ones defined by the distance function,

\begin{equation}
V_{\varepsilon, \Gamma} := \{(x,y)\in\field{R}{2}: d_{\Gamma}(x, y)\leq\varepsilon\}.
\end{equation} 

This is another reason why the study of a constraining potential given by the distance is interesting.

For technical reasons, much more attention has been devoted to the case of Neumann boundary conditions, which is by now well understood (Exner and Post (2005), Kuchment and Zeng (2001, 2003), Rubinstein and Schatzman (2001), Post (2006)). 

With the exception of Post (2006), all the papers we have quoted deal with the \emph{convergence of the spectrum} of the Laplacian, defined on a thickened graph, to the spectrum of a self-adjoint operator, defined on a graph whose edges have finite length, while we are primarily interested in the convergence of the dynamics and eventually to scattering theory.

A weak form of resolvent convergence for Neumann rigid approximations was studied by Sait\=o (Sait\=o (2000, 2001)), but his results do not allow to infer the structure of time evolution. 

In Post (2006) these results are improved, and norm resolvent convergence is established, under the hypotheses that the vertex neighbourhoods are ``small'' (for the precise meaning of the term we refer to the original paper) and, for graphs embedded in $\field{R}{2}$, that the angle between two different edges has a global lower bound. 

Another important difference is related to the class of initial conditions we consider. From a physical point of view, one expects that, when $\varepsilon \to 0$, the Hamiltonian \eqref{heps} gives rise to fast oscillations of the electron in the directions orthogonal to the edges of the graph. To prove the results mentioned above, one projects, roughly speaking, on the ground state of this transverse oscillation. Nonetheless, in the spirit of adiabatic perturbation theory (Teufel (2003)) we expect to be able to get an effective dynamics inside every transverse subspace, because they become broadly separated in energy when $\varepsilon$ goes to zero. For this reason, we consider initial wave functions which are localized inside one edge, and belong to an eigenspace of the transverse Hamiltonian (which will be defined more precisely in section II).

We consider longitudinal initial conditions which are independent of $\varepsilon$. This corresponds to study longitudinal states which vary over a wavelength which is much bigger than the transverse one. As it has been stressed in Belov \emph{et al.} (2004) however, longitudinal states are not homogeneous, and it would be interesting to consider also wavefunctions which vary on a scale of order $\varepsilon^{1/2}$ for example, analyzing in this way a semiclassical limit.

One should also remark that, in the Neumann case, the energy of the transverse ground state is independent of $\varepsilon$, because the Neumann Laplacian has always the eigenvalue zero corresponding to the constant function, while in the Dirichlet and soft approximation cases, the energy of every transverse mode tends to infinity when $\varepsilon\to 0$. This makes it necessary to subtract a divergent phase to get a finite result.

We will now describe briefly the structure of this paper. 

In section II we study the convergence of the unitary group generated by \eqref{heps} for an arbitrary graph (i. e., with an arbitrary number of vertices and edges). As we have already mentioned, we consider initial conditions which belong to a transverse eigenspace and are localized inside one edge of the graph and we choose a constraining potential given by the distance of a point from the graph. Using weak convergence methods we study the limit flow on the graph outside the vertices. To describe completely the limit flow, we must study its structure in a neighborhood of the vertices.

This may be a difficult task; this can be seen from the exact treatment we give in section III of a simpler system, in which the graph is a continuous curve in the plane. 

In this example, the graph has one vertex and two straight edges at an angle $0<\vartheta<\pi$. In this case we approximate the graph by a sequence of smooth curves converging to the graph when $\varepsilon\to 0$. We prove that generically (in particular if the curvature of the approximating curves is everywhere non-negative) the limit dynamics along the graph correspond to Dirichlet boundary conditions at the vertex. 

The proof is achieved by reducing the problem to the study of the dynamics with Dirichlet boundary conditions on the boundary of narrow tubes containing the graph, using a refined version of a theorem due to Froese and Herbst (2001). This however is not possible for every smooth curve approximating the graph, and a condition on the curvature comes in. From this result, it seems that the constraining potential and Dirichlet boundary conditions are not always interchangeable, as one could naively think.

Even though the geometry of the graph is very simple, this model demonstrates a mechanism through which adiabatic decoupling among different transverse modes takes place. In particular, it shows that the bound states localized near the vertices that can arise (and indeed do arise if instead of the constraining potential one considers a narrow tube with Dirichlet boundary conditions, see Carini \emph{et al.}(1992, 1993), Goldstone and Jaffe (1992)) do not interfere with the propagation of product states localized inside one of the edges \emph{at the leading order}, because their spectral distance becomes infinite in the limit.
\\ \\

\textbf{II. CONVERGENCE OUTSIDE THE VERTICES}

In this section we consider a finite metric graph, denoted by $\Gamma$, whose edges can have infinite length. We assume that it is embedded in $\field{R}{2}$ and, for the sake of simplicity, that all the edges are straight lines.

We denote by $V = \{v_{i}\}_{i\in I}$ the (finite) set of vertices and by $E = \{e_{j}\}_{j\in J}$ the (finite) set of edges connecting them. We assume that there are no isolated vertices.

A graph is said to be a metric graph if to each edge $e$ is assigned a length $l_{e}\in (0, +\infty]$. Edges of infinite length arise naturally if one considers scattering theory on graphs (see, e. g., Melnikov and Pavlov (1995) and references therein).

We can now identify each edge with a finite or infinite interval $[0, l_{e}]$, with the natural coordinate $x_{e}$ along it. One can also define function spaces (e. g., $L^{p}$ spaces, Sobolev spaces); in the case of Sobolev spaces, one must have some care at the vertices (see Kuchment (2004)).

As mentioned in the introduction, we approximate the dynamics on the graph using an Hamiltonian, acting on $L^{2}(\field{R}{2})$, with a constraining potential given by the square of the distance from $\Gamma$:
\begin{equation}\label{hamgenericgraph}\begin{split}
& \widehat{H}(\varepsilon) = -\frac{1}{2}\Delta + \frac{1}{2\varepsilon^{2}}d_{\Gamma}^{2},\\
& d_{\Gamma}(q) := \inf_{\tilde{q}\in\Gamma}\abs{q-\tilde{q}}, \qquad q, \tilde{q}\in\field{R}{2}.
\end{split}
\end{equation} 
One could use a different potential, whose Taylor expansion away from the vertices contains higher order terms, but we make the important assumption that the Hessian is constant along the edges with the same value on all edges. This condition is reasonable in view of the analysis in Dell'Antonio and Tenuta (2004) and Froese and Herbst (2001) when the system is constrained to smooth submanifolds. 

In this case, the energy of the transverse oscillation appears as a potential energy in the longitudinal motion, in the form $\omega(x)/\varepsilon$, where $\omega$ is the frequency of the oscillation. If $\omega$ does not depend on $x$, the resulting phase factor in the dynamics can be discarded. Otherwise, it originates a constraining potential along the edge, so that in the limit $\varepsilon\to 0$ we expect that the wave function concentrates along the minima of $\omega$. 

We denote by $\widehat{U}_{t}(\varepsilon)$ the unitary evolution associated to Hamiltonian \eqref{hamgenericgraph},
\begin{equation}
\widehat{U}_{t}(\varepsilon) := \exp(-it\widehat{H}(\varepsilon)),
\end{equation}
and we take as initial a state which ``lies in a subband'', i. e., is in a fixed transverse mode and localized within one edge. These are the states which are thought to describe the propagation of particles in semiconductor structures.

We have then
\begin{equation}\label{initialstate}
\psi_{0}(x_{e_{j_{0}}}, y_{e_{j_{0}}}) = f(x_{e_{j_{0}}})\Phi_{n}^{\varepsilon}(y_{e_{j_{0}}}),
\end{equation}
where $f\in\textrm{C}^{\infty}_{0}(0, l_{e_{j_{0}}})$ and $\Phi_{n}^{\varepsilon}$ is an eigenstate of the harmonic oscillator,
\begin{equation}\label{definitionofphin}\begin{split}
&\bigg(-\frac{1}{2}\frac{\partial^{2}}{\partial y^{2}} + \frac{1}{2\varepsilon^{2}}y^{2}\bigg)\Phi_{n}^{\varepsilon}(y) = \frac{E_{n}}{\varepsilon}\Phi_{n}^{\varepsilon}(y),\\
& E_{n} = n + \frac{1}{2}.
\end{split}
\end{equation}
$x_{e_{j_{0}}}$ is the natural coordinate along the edge $e_{j_{0}}$ and $y_{e_{j_{0}}}$ is the corresponding coordinate in the orthogonal direction. $x_{e_{j_{0}}}$ varies in the interval $[0, l_{e_{j_{0}}}]$ (or $[0, +\infty)$ if the edge has infinite length) and, since the edges are straight lines, $y_{e_{j_{0}}}$ is well defined and assumes values between $-\infty$ and $+\infty$ (to simplify the notation, from now we denote these coordinates just by $x_{j}$ and $y_{j}$).
\begin{figure}[h]\caption{Schematic representation of the initial state}
\begin{center}
\includegraphics[scale=0.4]{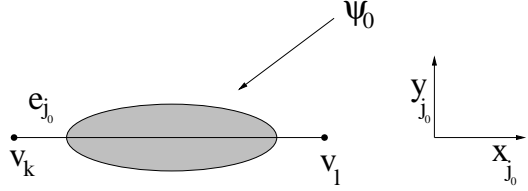}
\end{center}
\end{figure}

Applying $\widehat{U}_{t}(\varepsilon)$ to $\psi_{0}$ we expect the appearance of a strongly oscillating factor, given by $\exp(-iE_{n}t/\varepsilon)$. To avoid this (irrelevant) phase, we consider the modified unitary group:
\begin{equation}\begin{split}
& \widetilde{H}(\varepsilon) := \widehat{H}(\varepsilon) - \frac{E_{n}}{\varepsilon},\\
& \psi_{t}^{\varepsilon} := \widetilde{U}(\varepsilon)\psi_{0} := \exp(-it\widetilde{H}(\varepsilon))\psi_{0}.
\end{split}
\end{equation}

To analyze the adiabatic decoupling, we split $\psi_{t}^{\varepsilon}$ according to the different transverse components in each edge:
\begin{equation}
s_{j}^{m}(t, x_{j};\varepsilon) := \int\ dy_{j}\ \Phi_{m}^{\varepsilon}(y_{j})^{*}\psi_{t}^{\varepsilon}(x_{j}, y_{j})\footnote[1]{With an abuse of notation, we denote by $\psi_{t}^{\varepsilon}(x_{j}, y_{j})$ the function $\psi_{t}^{\varepsilon}$ written in coordinates $(x_{j}, y_{j})$. Since the different systems of coordinates associated to each edge are linked to one another by a rigid motion of the plane, this does not modify the differentiability or integrability properties of $\psi_{t}^{\varepsilon}$. Note, instead, that $\Phi_{m}^{\varepsilon}(y_{j})$ is an eigenfunction of the harmonic oscillator in the variable $y_{j}$.},
\end{equation}
where $x_{j}$ is the natural coordinate along $e_{j}$ and $y_{j}$ is orthogonal to it. 

\begin{prop}\label{projectionbounded} Let $P_{j}^{m}$ the operator from $\mathcal{S}(\field{R}{2})$ to $H^{k}(\field{R}{})$ defined by
\begin{equation}
P_{j}^{m}\psi(x_{j}) := \int_{\field{R}{}}\, dy_{j} \, \Phi_{m}^{\varepsilon}(y_{j})^{*}\psi(x_{j}, y_{j}),  
\end{equation}
then $P_{j}^{m}$ extends to a unique operator (of norm $1$) from $H^{k}(\field{R}{2})$ to $H^{k}(\field{R}{})$, for every $k\in\field{N}{}$, $k\geq 0$.
\end{prop}

\begin{proof}
Given $\psi\in\mathcal{S}(\field{R}{2})$ it is clear that
\begin{equation*}\begin{split}
&\deriv{x}{l} (P_{j}^{m}\psi)(x) = \int_{\field{R}{}}\ dy \ \Phi_{m}(y)^{*}\deriv{x}{l}\psi(x, y) = <\Phi_{m},
\deriv{x}{l}\psi>_{L^{2}(\field{R}{}_{y})}\\
& \Rightarrow |\deriv{x}{l} (P_{j}^{m}\psi)(x)|^{2}\leq \norm{\Phi_{m}}^{2}_{L^{2}(\field{R}{})}\cdot\int\ dy |\deriv{x}{l}\psi(x, y)|^{2} \\
& \Rightarrow \int\ dx |\deriv{x}{l} (P_{j}^{m}\psi)(x)|^{2} \leq \norm{\Phi_{m}}^{2}_{L^{2}(\field{R}{})} \cdot
\norm{\deriv{x}{l}\psi}^{2}_{L^{2}(\field{R}{2})}
\end{split}
\end{equation*}
\begin{equation}
 \Rightarrow  \norm{P_{j}^{m}\psi}_{H^{k}(\field{R}{})}^{2} = \sum_{l=0}^{k} \norm{\deriv{x}{l}P_{j}^{m}\psi}_{L^{2}(\field{R}{})}^{2}\leq \norm{\Phi_{m}}^{2}\sum_{l=0}^{k}
\norm{\deriv{x}{l}\psi(x, y)}_{L^{2}(\field{R}{2})}^{2}.
\end{equation}
\end{proof}

\begin{cor}\label{sjbounded}
The components $s_{j}^{m}(t, x_{j}; \varepsilon)$ are well defined, belong to $H^{1}(\field{R}{})$ in the variable $x_{j}$ and satisfy
\begin{equation}
\sup_{t\in[0, T]}\norm{s_{j}^{m}(t, \cdot; \varepsilon)}_{L^{2}(\field{R}{})}\leq \textrm{const.}\qquad .
\end{equation}
\end{cor}
\begin{proof}
The domain of the quadratic form associated to $\widetilde{H}(\varepsilon)$ is given by
\begin{equation}
Q(\widetilde{H}(\varepsilon)) := H^{1}(\field{R}{2}) \cap Q(d^2) \quad ,
\end{equation}
where $Q(d^2) := \{\psi \in L^{2}(\field{R}{2}): d_{\Gamma}(x,y)\psi \in L^{2}(\field{R}{2})\}$. $\psi_{0}$ belongs to $Q(\widetilde{H}(\varepsilon))$, so $\psi_{t}^{\varepsilon}$ is in $H^{1}(\field{R}{2})$.
\end{proof}

\begin{lem}\label{hlim}
\begin{equation}
\norm{\widetilde{H}(\varepsilon)\psi_{0}}\leq C\quad (\textrm{independent of}\quad\varepsilon).
\end{equation}
\end{lem}
\begin{proof}
Since the Laplacian is invariant by rotations and translations, we have (for simplicity we drop the index $j_{0}$ in $x$ and $y$)
\begin{equation*}\begin{split}
\widetilde{H}(\varepsilon)\psi_{0} & =  -\frac{1}{2}\deriv{x}{2}f\cdot \Phi_{n} - \frac{1}{2}f\cdot\deriv{y}{2}\Phi_{n} +
\frac{1}{2\varepsilon^{2}}d_{\Gamma}^{2}f\cdot \Phi_{n} - \frac{E_{n}}{\varepsilon}f\cdot \Phi_{n} = \\
& = -\frac{1}{2}\deriv{x}{2}f\cdot \Phi_{n} + \frac{1}{2\varepsilon^{2}}(d_{\Gamma}^{2} - y^{2})f\cdot \Phi_{n} -
\frac{1}{2}f\cdot\deriv{y}{2}\Phi_{n} +\\
&+ f\frac{1}{2\varepsilon^{2}}y^{2}\cdot \Phi_{n} - \frac{E_{n}}{\varepsilon}f\cdot \Phi_{n}=\\
& = -\frac{1}{2}\deriv{x}{2}f\cdot \Phi_{n} + \frac{1}{2\varepsilon^{2}}(d_{\Gamma}^{2} - y^{2})f\cdot \Phi_{n}.
\end{split}
\end{equation*}

Let $a$ be the infimum of the support of $f$, and $b>a$ the supremum. Since $f$ is supported inside the edge $e_{j_{0}}$ and near each edge the distance from the graph is equal to $\abs{y_{j}}$, the function $\frac{1}{2\varepsilon^{2}}(d_{\Gamma}^{2} - y^{2})f\cdot \Phi_{n}$ will be zero when $\abs{y} < D$, where $D$, depending on the support of $f$, can be small, but it is strictly positive. Therefore we have
\begin{equation}\label{norm}
\norm{\frac{1}{2\varepsilon^{2}}(d_{\Gamma}^{2} - y^{2})f\cdot \Phi_{n}}^{2} = \int_{a}^{b}\ dx \int_{\abs{y}>D}\ dy\ \frac{1}{4\varepsilon^{4}}
(d_{\Gamma}^{2} - y^{2})^{2}|f(x)\Phi_{n}^{\varepsilon}(y)|^{2} .
\end{equation}
Now we use the following two properties:
\begin{itemize}
\item $d_{\Gamma}^{2}$ is equal to a polynomial of second order in the variables $(x, y)$;
\item $\Phi^{\varepsilon}_{n}$ is equal to a polynomial in $y/\varepsilon^{1/2}$ times $\exp(-y^{2}/2\varepsilon)$. 
\end{itemize}
The norm \eqref{norm} contains then terms of the form ($P$ and $Q$ are polynomials)
\begin{equation*}\begin{split}
& \int_{a}^{b}\ dx\ \abs{f(x)}^{2} \int_{\abs{y}>D}\ dy\ P(x, y)Q(x, y/\varepsilon^{1/2})\exp\bigg(-\frac{y^{2}}{\varepsilon}\bigg) =\\
&= \int_{a}^{b}\ dx\ \abs{f(x)}^{2} \int_{\abs{y}>D}\ dy\ P(x, y)Q(x, y/\varepsilon^{1/2})\exp\bigg[-\bigg(\frac{1}{\varepsilon} - M\bigg) y^{2}\bigg]\exp(-My^{2})\leq\\
&\leq \exp\bigg[-\bigg(\frac{1}{\varepsilon} - M\bigg) D^{2}\bigg]\int_{a}^{b}\ dx\ \abs{f(x)}^{2} \int_{\abs{y}>D}\ dy\ P(x, y)Q(x, y/\varepsilon^{1/2})\exp(-My^{2})=\\
&= O(e^{-c/\varepsilon}).
\end{split}  
\end{equation*}
This implies that
\begin{equation}\label{expsmall}
\norm{\frac{1}{2\varepsilon^{2}}(d_{\Gamma}^{2} - y^{2})f\cdot \Phi_{n}}^{2} = O(e^{-c/\varepsilon}),
\end{equation}
and therefore the thesis is proved.
\end{proof}

\begin{cor}
For every system of coordinates $(x_{j}, y_{j})$ associated to an edge $e_{j}$ we have
\begin{equation}\label{energyestimate}
\frac{1}{2}\norm{\deriv{x_{j}}{}\psi_{t}^{\varepsilon}}^{2} + 
\frac{1}{2}\norm{\deriv{y_{j}}{}\psi_{t}^{\varepsilon}}^{2} + 
 \frac{1}{2\varepsilon^{2}}\norm{d_{\Gamma} \psi_{t}^{\varepsilon}}^{2} \leq \frac{C}{\varepsilon}.
\end{equation}
\end{cor}

\begin{cor}\label{zerofar}
Let $F_{d_{\Gamma}\geq\delta}$ be the characteristic function of the set $\{(x, y): d_{\Gamma}(x, y)\geq\delta\}$, where $\delta$ is any positive number, then
\begin{equation}
\sup_{t\in[0, T]} \norm{F_{d_{\Gamma}\geq\delta}\ \psi_{t}^{\varepsilon}}_{L^{2}(\field{R}{2})} = O(\varepsilon^{1/2}).
\end{equation}
\begin{proof}
Last term in \eqref{energyestimate} gives 
\begin{equation*}
\frac{1}{2}\norm{d_{\Gamma}\psi_{t}^{\varepsilon}}^{2} \leq C\varepsilon.
\end{equation*}

Therefore we get
\begin{equation*}
\delta^{2}\ \cdot \ <\psi_{t}^{\varepsilon},\ F_{d_{\Gamma}\geq\delta}\ \psi_{t}^{\varepsilon}>
 \leq \  <d_{\Gamma}\ \psi_{t}^{\varepsilon},\ F_{d_{\Gamma}\geq\delta}\ d_{\Gamma}\ \psi_{t}^{\varepsilon}> \leq \ 2C\varepsilon.
\end{equation*}
\end{proof}
\end{cor}

We are now ready to prove 
\begin{theo}\label{adiabaticseparation}
For $m \neq n$ and $j\in J$ 
\begin{equation}
s_{j}^{m}(t, x_{j}; \varepsilon) \weakstar 0, \quad \varepsilon \to 0,
\end{equation}
where the convergence is in the weak$^{*}$ topology of $L^{\infty}((0, T), L^{2}(\field{R}{}))$.
\end{theo}
\begin{proof}
For convenience of the reader, we recall that a \emph{bounded} sequence $f_{\varepsilon}$ of functions in $L^{\infty}((0, T), L^{2}(\field{R}{}))$ converges to a limit $f_{0}\in L^{\infty}((0, T), L^{2}(\field{R}{}))$ in the weak$^{*}$ topology if and only if
\begin{equation*}
\int_{0}^{T}\ dt\ \varphi(t) <\chi(\cdot), f_{\varepsilon}(t, \cdot)>_{L^{2}(\field{R}{})} \to \int_{0}^{T}\ dt\ \varphi(t) <\chi(\cdot), f_{0}(t, \cdot)>_{L^{2}(\field{R}{})} 
\end{equation*}
for every function $\varphi\in L^{1}(0, T)$ and every function $\chi\in L^{2}(\field{R}{})$. 

It is a standard fact about weak$^{*}$ topology that it is enough to consider only $\varphi$ and $\chi$ in dense subsets of $L^{1}((0, T))$ and $L^{2}(\field{R}{})$ respectively (see, e. g., Bornemann (1998), Appendix B or Rudin (1973), theorems $3.15$ and $3.16$).

Let us then show first that
\begin{equation}
\sup_{t\in [0, T]}\ \abs{<\chi(\cdot), s_{j}^{m}(t, \cdot; \varepsilon)>_{L^{2}(\field{R}{})}} \to 0, \quad \varepsilon\to 0, 
\end{equation}
for every function $\chi\in \textrm{C}^{\infty}_{0}(\field{R}{}\backslash\{0, l_{e_{j}}\})$ if the edge $e_{j}$ has finite length and for every $\chi\in \textrm{C}^{\infty}_{0}(\field{R}{}\backslash\{0\})$ if $e_{j}$ has infinite length. We consider explicitly only the latter case, the former being analogous.

If the support of $\chi$ is contained in $(-\infty, 0)$ then
\begin{equation*}\begin{split}
& \abs{<\chi, s_{j}^{m}>_{L^{2}(\field{R}{})}} = \abs{<\chi\cdot \Phi_{m}^{\varepsilon}, \psi_{t}^{\varepsilon}>_{L^{2}(\field{R}{2})}} = \abs{<\chi\cdot \Phi_{m}^{\varepsilon}, F_{\chi}\ \psi_{t}^{\varepsilon}>} \\
&\leq \norm{\chi\cdot \Phi_{m}^{\varepsilon}}\cdot
\norm{F_{\chi}\ \psi_{t}^{\varepsilon}} = O(\varepsilon^{1/2}),
\end{split}
\end{equation*}
where $F_{\chi}$ is the characteristic function of the support of $\chi$ and we have used corollary \ref{zerofar}.

If the support of $\chi$ in contained in $(0, +\infty)$, then, following the proof of lemma \ref{hlim} we can show that
\begin{equation*}\begin{split}
\widetilde{H}(\varepsilon)[\chi(x_{j})\Phi_{m}^{\varepsilon}(y_{j})] =&\ \textrm{(we drop the index}\ j) \quad -\frac{1}{2}\deriv{x}{2}\chi\cdot \Phi_{m}^{\varepsilon} + \\
&+ \frac{1}{2\varepsilon^{2}}(d_{\Gamma}^{2} - y^{2})\chi\cdot \Phi_{m}^{\varepsilon} + \frac{E_{m}-E_{n}}{\varepsilon}\chi\cdot\Phi_{m}^{\varepsilon}.
\end{split}
\end{equation*}
Since $\chi$ is supported away from the vertex (located at $x_{j}=0$), an equation similar to \eqref{expsmall} holds:
\begin{equation*}
\norm{\frac{1}{2\varepsilon^{2}}(d_{\Gamma}^{2} - y^{2})\chi\cdot \Phi_{m}^{\varepsilon}}^{2} = O(e^{-c/\varepsilon}).
\end{equation*}

Since $m\neq n$, we have then
\begin{equation*}
\chi\Phi_{m}^{\varepsilon} = \frac{\varepsilon}{m-n}\bigg\{\widetilde{H}(\varepsilon)[\chi(x_{j})\Phi_{m}^{\varepsilon}(y_{j})] + \frac{1}{2}\deriv{x}{2}\chi\cdot \Phi_{m}^{\varepsilon}\bigg\} + O(e^{-c/\varepsilon}).
\end{equation*}
This implies
\begin{equation*}\begin{split}
<\chi(\cdot), s_{j}^{m}(t, \cdot; \varepsilon)> = <\chi(\cdot)\Phi_{m}^{\varepsilon}, \psi_{t}^{\varepsilon}> &= \frac{\varepsilon}{m-n}<\chi(\cdot)\Phi_{m}^{\varepsilon}, \widetilde{H}(\varepsilon)\psi_{t}^{\varepsilon}> + \\
&+ \frac{\varepsilon}{m-n}<\frac{1}{2}\deriv{x}{2}\chi\cdot \Phi_{m}^{\varepsilon}, \psi_{t}^{\varepsilon}> + O(e^{-c/\varepsilon}) =\\
&=\ (\textrm{lemma \ref{hlim}})\ O(\varepsilon).
\end{split}
\end{equation*} 
Now, if $\varphi\in L^{1}((0, T))$, we get
\begin{equation*}
\bigg\lvert\int_{0}^{T}\ dt\ \varphi(t) <\chi(\cdot), s_{j}^{m}(t, \cdot; \varepsilon)>_{L^{2}(\field{R}{})}\bigg\rvert \leq  \norm{\varphi}_{L^{1}} \cdot
\sup_{t\in [0, T]}\ |<\chi(\cdot), s_{j}^{m}(t, \cdot; \varepsilon)>|,
\end{equation*}
but we have just shown that the right-hand side goes to zero for $\chi\in\textrm{C}^{\infty}_{0}(\field{R}{}\backslash\{0\})$ (or $\textrm{C}^{\infty}_{0}(\field{R}{}\backslash\{0, l_{e_{j}}\})$ for an edge of finite length) which is dense in $L^{2}(\field{R}{})$.
\end{proof}

Theorem \ref{adiabaticseparation} shows that, although in a weak sense, there is indeed adiabatic separation between the different transverse states even in the presence of vertices, if the initial state is localized in two senses: first, it has to be localized inside one edge to avoid mixing between the different transverse states associated to each edge and second, it has to be in one (or a finite number of) transverse band(s).

Since the limit of $s_{j}^{m}$ for $m\neq n$ is zero, to analyze in a complete way the (limit) evolution of $\psi_{0}$ we have to determine the behaviour of $s_{j}^{n}$, $j\in J$ as a function of time.

\begin{theo}\label{convergenceoutside}
There exists a weak$^{*}$ convergent subsequence of $s_{j}^{n}(t, x_{j}; \varepsilon)$ in $L^{\infty}((0, T), L^{2}(\field{R}{}))$ (denoted again by the same symbol), whose limit $s_{j}^{n}(t, x_{j}; 0)\in L^{\infty}((0, T), L^{2}(\field{R}{}))$ satisfies
\begin{equation}
i\deriv{t}{}s_{j}^{n}(t, x_{j}; 0) = -\frac{1}{2}\deriv{x}{2}s_{j}^{n}(t, x_{j}; 0)\quad \textrm{in}\quad \mathcal{D}'((0, T)\times(0, l_{e_{j}})).
\end{equation}
\end{theo}

\begin{rem}
By corollary \ref{sjbounded}, $s_{j}^{n}(t, x_{j}; \varepsilon)$ is a bounded sequence in $L^{\infty}((0, T), L^{2}(\field{R}{}))$. Since the balls in $L^{\infty}((0, T), L^{2}(\field{R}{}))$ are compact metric spaces with respect to the weak$^{*}$ topology (see the theorems in the book of Rudin quoted above), a weak$^{*}$ convergent subsequence certainly exists. 

Moreover, if one shows that all the weak$^{*}$ convergent subsequences converge to the same limit, then this implies that the sequence itself converges. 

The equation satisfied by the limit in theorem \ref{convergenceoutside} is clearly independent of the subsequence, but it does not determine the behaviour of the limit in the vertices, so we cannot conclude convergence of the sequence. 

For this it would be necessary to control the behavior of the sequence in a neighborhood of the vertices. This difficulty (which is not present for smooth submanifolds, where the same strategy has been successfully applied by Bornemann (1998) in the classical case) is linked with the fact that the operator $-d^{2}/dx^{2}$ defined for functions which vanish in a neighborhood of the origin has many self-adjoint extensions which define different dynamics.
\end{rem}

We split the proof of the theorem into a number of lemmas.

\begin{lem}
$s_{j}^{n}(t, x_{j}; \varepsilon)$ belongs to \emph{C}$^{1}([0, T], L^{2}(\field{R}{}))$ and moreover it is an equicontinuous sequence of function from $[0, T]$ to $L^{2}(\field{R}{})$.  
\end{lem}

\begin{proof}
Let us denote by $\tilde{s}_{j}^{n}(t, x_{j}; \varepsilon) \in \textrm{C}^{0}([0, T], L^{2}(\field{R}{}))$ the function
\begin{equation*}
P_{j}^{n}[-i\widetilde{H}(\varepsilon)\psi_{t}^{\varepsilon}].
\end{equation*}

Using proposition \ref{projectionbounded} we have that
\begin{equation*}\begin{split}
&\bigg\lVert \frac{s_{j}^{n}(t+h, \cdot; \varepsilon) - s_{j}^{n}(t, \cdot; \varepsilon)}{h} - \tilde{s}^{n}_{j}(t, \cdot; \varepsilon)\bigg\rVert_{L^{2}(\field{R}{})} = \bigg\lVert P_{j}^{n}\bigg[\frac{\widetilde{U}_{t+h}(\varepsilon)-\widetilde{U}_{t}(\varepsilon)}{h} + i\widetilde{U}_{t}(\varepsilon)\widetilde{H}(\varepsilon)\bigg]\psi_{0} \bigg\rVert \\
&\leq \bigg\lVert\bigg[\frac{\widetilde{U}_{t+h}(\varepsilon)-\widetilde{U}_{t}(\varepsilon)}{h} + i\widetilde{U}_{t}(\varepsilon)\widetilde{H}(\varepsilon)\bigg]\psi_{0}\bigg\rVert \to 0.
\end{split}
\end{equation*}
This proves that 
\begin{equation}\label{derivlim}
i\deriv{t}{}s_{j}^{n}(t, x_{j}; \varepsilon) = \tilde{s}_{j}^{n}(t, x_{j}; \varepsilon) = P_{j}^{m}[-i\widetilde{U}_{t}(\varepsilon)\widetilde{H}(\varepsilon)\psi_{0}].
\end{equation}

Since $\norm{\widetilde{H}(\varepsilon)\psi_{0}}$ is bounded (lemma \ref{hlim}), $\norm{\tilde{s}_{j}^{m}(t, x_{j}; \varepsilon)}$ is bounded, therefore
\begin{equation*}\begin{split}
&\norm{s_{j}^{m}(t, \cdot; \varepsilon) - s_{j}^{m}(t', \cdot; \varepsilon)}_{L^{2}(\field{R}{})} = \bigg\lVert \int_{t}^{t'} d\tau\ 
\deriv{\tau}{}s_{j}^{m}(\tau, \cdot; \varepsilon) \bigg\rVert \\
& \leq  \int_{t}^{t'} d\tau\ \norm{\deriv{\tau}{}s_{j}^{m}(\tau, \cdot; \varepsilon)} \leq C\abs{t-t'},
\end{split}
\end{equation*}
showing that $s_{j}^{m}(t, x_{j}; \varepsilon)$ is an equicontinuous sequence.
\end{proof}

\begin{cor}\label{propertieslimit}
There exists a subsequence $s_{j}^{m}(t, x_{j}; \varepsilon)$ which satisfies:
\begin{enumerate}
\item $s_{j}^{m}(t, x_{j}; \varepsilon)$ converges, in the weak topology of $L^{2}(\field{R}{})$, uniformly in $t$, to a limit $s_{j}^{m}(t, x_{j}; 0)\in L^{2}(\field{R}{})$. Moreover, the limit is continuous in $t$ in the weak topology of $L^{2}$.
\item $\deriv{t}{}s_{j}^{m}(t, x_{j}; \varepsilon) \weakstar \deriv{t}{}s_{j}^{m}(t, x_{j}; 0)\ \textrm{in}\ L^{\infty}(\ (0, T), L^{2}(\field{R}{}))$, where the derivative $\deriv{t}{}s_{j}^{m}(t, x_{j}; 0)$ is to be interpreted as derivative in $\mathcal{D}'((0, T)\times\field{R}{}_{x})$.
\end{enumerate}
\end{cor}
\begin{proof}
The sequence $s_{j}^{m}(t, x_{j}; \varepsilon)$ is contained in a ball in $L^{2}(\field{R}{})$. This ball is a compact metric space with respect to the weak topology. Since the sequence is equicontinuous with respect to the strong topology, it will be equicontinuous with respect to the weak topology too. Therefore, the theorem of Ascoli-Arzel\`a (see, e. g., Royden (1988), theorem $10.40$) proves the first point.

Equation \eqref{derivlim} implies that $\deriv{t}{}s_{j}^{m}(t, x_{j}; \varepsilon)$ is a bounded sequence  in $L^{\infty}(\ (0, T), L^{2}(\field{R}{}))$, so, extracting possibly another subsequence, we have that there exists $g_{j}^{m}\in L^{\infty}(\ (0, T), L^{2}(\field{R}{}))$ such that (again, we denote the subsequence with the same symbol as the sequence itself) 
\begin{equation*}
\deriv{t}{}s_{j}^{m}(t, x_{j}; \varepsilon) \weakstar g_{j}^{m},
\end{equation*}
but this implies that $\forall$ $\varphi\in\textrm{C}^{\infty}_{0}((0, T))$, $\forall$ $\chi\in\textrm{C}^{\infty}_{0}(\field{R}{})$,
\begin{equation*}\begin{split}
&\int_{0}^{T}dt\int_{\field{R}{}}dx\ g_{j}^{m}(t, x)\varphi(t)\chi(x) =\\
& = \int_{0}^{T}dt\ \varphi(t) <\chi(\cdot), g_{j}^{m}(t, \cdot)>_{L^{2}(\field{R}{})} \gets \int_{0}^{T}dt\ \varphi(t) <\chi, 
\deriv{t}{}s_{j}^{m}(t, \cdot; \varepsilon)> = \\
& = \int_{0}^{T}dt\ \varphi(t) \deriv{t}{}<\chi, s_{j}^{m}(t, \cdot; \varepsilon)> =\\
&= -\int_{0}^{T}dt\ \deriv{t}{}\varphi <\chi, s_{j}^{m}(t, \cdot; \varepsilon)> \to 
-\int_{0}^{T}dt\ \deriv{t}{}\varphi <\chi, s_{j}^{m}(t, \cdot; 0)> =\\
&= \int_{0}^{T}dt\int_{\field{R}{}}dx\ s_{j}^{m}(t, x; 0)\deriv{t}{}\varphi(t)\chi(x),
\end{split}
\end{equation*}
\begin{equation*}
\Rightarrow g_{j}^{m} = \deriv{t}{}s_{j}^{m}(t, x_{j}; 0)\ \textrm{in}\ \mathcal{D}'((0, T)\times\field{R}{}_{x}).
\end{equation*}
\end{proof}

We can now prove theorem \ref{convergenceoutside}.
\begin{proof} 
We suppose that edge $e_{j}$ has infinite length. The proof for an edge of finite length is analogous.

Corollary \ref{zerofar}, together with the proof of the first part of the proof of theorem \ref{adiabaticseparation} implies that
\begin{equation*}
\sup_{t\in [0, T]}\ |<\chi, s_{j}^{m}(t, \cdot; \varepsilon)>| = O(\varepsilon^{1/2}),
\end{equation*}
for all $j\in J$ and for all $\chi\in\textrm{C}^{\infty}_{0}(-\infty, 0)$, but the first point of corollary \ref{propertieslimit} gives
\begin{equation*}
<\chi, s_{j}^{m}(t, \cdot; 0)>_{L^{2}(\field{R}{})} = \lim_{\varepsilon\to 0} <\chi, s_{j}^{m}(t, \cdot; \varepsilon)> = 0.  
\end{equation*}
Equation \eqref{derivlim} allows us to write, for all $\chi\in\textrm{C}^{\infty}_{0}(0, +\infty)$,
\begin{equation*}\begin{split}
&<\chi, i\deriv{t}{}s_{j}^{m}(t, \cdot; \varepsilon)>_{L^{2}(\field{R}{})} = <\chi\cdot \Phi_{n}^{\varepsilon}, \widetilde{H}(\varepsilon)\psi_{t}^{\varepsilon}>_{L^{2}(\field{R}{2})} = <\widetilde{H}(\varepsilon) \chi\cdot \Phi_{n}^{\varepsilon}, \psi_{t}^{\varepsilon}> = \\
& = <-\frac{1}{2}\deriv{x}{2}\chi\cdot \Phi_{n}^{\varepsilon}, \psi_{t}^{\varepsilon}> + <\frac{1}{2\varepsilon^{2}}(d_{\Gamma}^{2} - y^{2}) \chi\cdot \Phi_{n}^{\varepsilon}, \psi_{t}^{\varepsilon}> = <-\frac{1}{2}\deriv{x}{2}\chi, s_{j}^{m}(t, \cdot; \varepsilon)>_{L^{2}(\field{R}{})} + \\
&+<\frac{1}{2\varepsilon^{2}}(d_{\Gamma}^{2} - y^{2}) \chi\cdot \Phi_{n}^{\varepsilon}, \psi_{t}^{\varepsilon}>
\end{split}
\end{equation*}

Since $\chi$ is supported in $(0, +\infty)$, equation \eqref{expsmall} holds also in this case, therefore
\begin{equation*}
\norm{\frac{1}{2\varepsilon^{2}}(d_{\Gamma}^{2} - y^{2})\chi\cdot \Phi_{n}^{\varepsilon}}^{2} = O(e^{-c/\varepsilon}).
\end{equation*}

We have then, for all $\varphi\in\textrm{C}^{\infty}_{0}(0, T)$, and for all $\chi\in\textrm{C}^{\infty}_{0}(0, +\infty)$,
\begin{equation*}\begin{split}
&\int_{0}^{T}dt\ \varphi(t)<\chi, i\deriv{t}{}s_{j}^{n}(t, \cdot; 0))>_{L^{2}(\field{R}{})} \gets 
\int_{0}^{T}dt\ \varphi(t)<\chi, i\deriv{t}{}s_{j}^{n}(t, \cdot; \varepsilon)>_{L^{2}(\field{R}{})} =\\ &=\int_{0}^{T}dt\ \varphi(t) <-\frac{1}{2}\deriv{x}{2}\chi, s_{j}^{n}(t, \cdot; \varepsilon)>_{L^{2}(\field{R}{})} +
 O(e^{-c/\varepsilon})\norm{\varphi}_{L^{1}(0, T)}\\
& \to \int_{0}^{T}dt\ \varphi(t) <-\frac{1}{2}\deriv{x}{2}\chi, s_{j}^{n}(t, \cdot; 0)>_{L^{2}(\field{R}{})},
\end{split}
\end{equation*}
\begin{equation}
\Rightarrow i\deriv{t}{}s_{j}^{n}(t, x_{j}; 0) = -\frac{1}{2}\deriv{x}{2}s_{j}^{n}(t, x_{j}; 0)\ \textrm{in}\ 
\mathcal{D}'((0, T)\times (0, +\infty)).
\end{equation}
\end{proof}

\textbf{III. A GRAPH WITH TWO EDGES}

In this section, we are going to put forward a different kind of soft approximation for a graph with one vertex and two infinite edges. We denote it by $\Gamma_{\vartheta}$, where $\vartheta$ is the angle made by the two edges, $0<\vartheta<\pi$.

As we have already said in the introduction, we do not consider directly $\gammatheta$, but we approximate it by smooth curves, $\Gamma_{\vartheta, \delta}$, whose curvature becomes bigger and bigger in a region whose width, given by $\delta$, goes to zero and we consider a potential constraining to this family of curves.

More precisely, to specify the approximating curves we need only to specify their curvature, $k_{\delta}$, because, as it is well known, this determines the curve up to rigid motions of the plane. Naturally, we want that, when $\delta$ goes to zero, the curves tend to the graph. This in particular implies that the turning angle has to become equal to $\vartheta$ when $\delta\to 0$.

A simple choice which satisfies these requests is ($s$ is the arc length parameter)
\begin{equation}\label{deltacurvature}\begin{split}
& k_{\delta}(s) := \frac{\vartheta}{\delta}k\bigg(\frac{s}{\delta}\bigg),\quad \int_{\field{R}{}}\ ds\ k(s)=1,\\
& k\in \textrm{C}^{\infty}_{0}(-1, 1),\quad \begin{cases}
k=1 & \abs{s}<1/2 \\
k=0 & \abs{s}>3/4,
                                                                                  \end{cases}
\end{split}
\end{equation}
which amounts to deformate the graph in a neighbourhood of the vertex replacing it with an arc of a circle. Note that the $\delta$ scaling is fixed by the request that the turning angle of the approximating curves be $\vartheta$,
\begin{equation*}
\int_{\field{R}{}}\ ds\ \frac{\vartheta}{\delta}k\bigg(\frac{s}{\delta}\bigg) = \vartheta.
\end{equation*}

Actually, our result does not depend on this specific choice we have made, because, from the proof, one can see that the only essential ingredient is the singularity $1/\delta$, which is forced by the requirement that the turning angle is $\vartheta$.

We consider the Hamiltonian
\begin{equation*}
\widehat{H}(\varepsilon, \delta(\varepsilon)) = -\frac{1}{2}\Delta + \frac{1}{\varepsilon^{2}}W_{\delta(\varepsilon)}, \quad \delta(\varepsilon)\to 0\ \textrm{when}\ \varepsilon\to 0,
\end{equation*}
where, for simplicity, we suppose that
\begin{equation*}
W_{\delta(\varepsilon)}(x, y)=\frac{1}{2}d_{\delta(\varepsilon)}^{2}(x, y) =\frac{1}{2}\textrm{dist}[(x, y), \Gamma_{\vartheta, \delta(\varepsilon)}]^{2}.
\end{equation*}

The remark we made above about the possibility to generalize the analysis to potentials with constant Hessian, applies here too. As in the previous section, we are interested in the time evolution of a product state which is initially localized away from the vertex.

We expect that the particle oscillates very fast along the direction normal to the curve, so to analyze the motion we should use a suitable system of coordinates, adapted to the curve. A natural choice is given by \emph{tubular coordinates}, which are a set of local coordinates suited to study tubular neighbourhoods of embedded submanifolds (see, e. g., Lang (1995)). If the submanifold one considers has codimension (and dimension) bigger than one, then the metric in tubular coordinates is in general not diagonal, and the Laplacian in these coordinates contains a gauge term which couples the longitudinal and the transverse motion (Mitchell (2001) and references therein).

In our case, since both the dimension and the codimension are equal to one, these problems do not appear.

Given a smooth curve $C$, with parametric equation $\zeta: \Omega\to\field{R}{2}$ such that $\abs{\deriv{s}{}\zeta(s)}=1$, we can describe the position of points in a tubular neighbourhood $N$ of $C$ via the curvilinear coordinates $(s, u)$ defined by
\begin{equation}
q(s, u) = \zeta(s) + u\boldsymbol{n}(s),
\end{equation} 
where $q$ is an arbitrary point of $N$, $\boldsymbol{n}(s)$ is the normal unit vector to the curve and $u$ is assumed to be smaller than the radius of curvature.

\begin{figure}
\begin{center}
\includegraphics[scale=0.4]{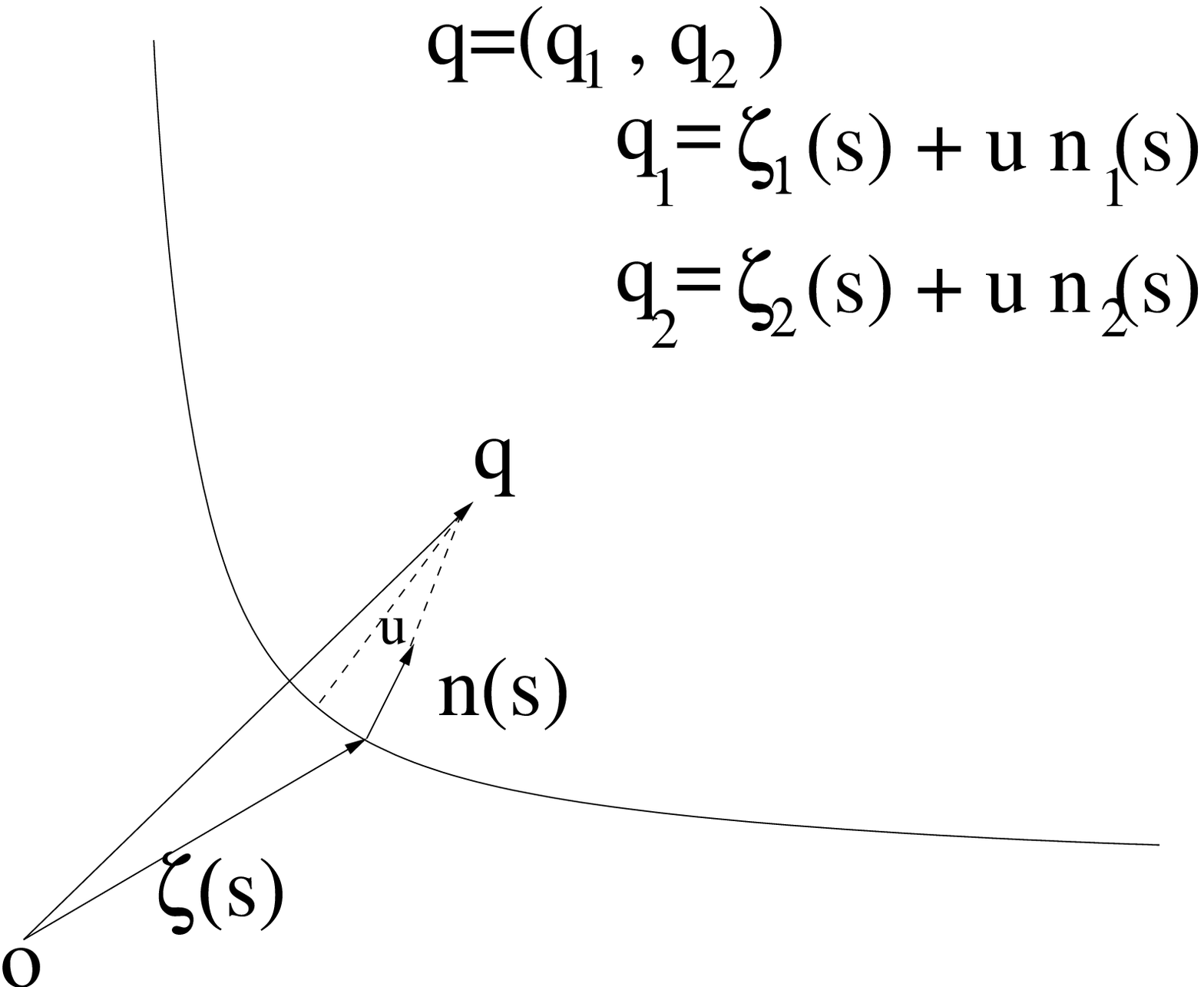}
\end{center}
\end{figure}

In the case we are dealing with, this means that curvilinear coordinates are defined only in the region 
\begin{equation}
\{(x, y)\in\field{R}{2}: d_{\delta}(x, y)<\varrho\},
\end{equation} 
where $\varrho$ is the radius of curvature of $\Gamma_{\vartheta, \delta}$. When $k_{\delta}$ is different from zero, this quantity is proportional to $\delta$ itself, so, by hypothesis, it goes to zero when $\varepsilon\to 0$.

To get rid of the region $\{(x, y): d_{\delta}(x, y)>\delta\}$ we will use a theorem, proved first by Froese and Herbst in the more general context of a potential constraining to a submanifold (proposition $8.1$ in Froese and Herbst (2001)), which basically says that if one starts from an initial state more and more localized near the constraint, then all that matters for the time evolution is a small region near the constraint itself. We repeat the proof of Froese and Herbst because we need to keep track of the dependence of all the constants in the estimates on $\delta$, to apply them to the region $\{(x, y)\in\field{R}{2}: d_{\delta}(x, y)<\delta(\varepsilon)\}$. 

\begin{theo}
Let $\psi\in L^{2}(\field{R}{2})$, $\norm{\psi}=1$ and $\norm{\widehat{H}(\varepsilon, \delta)\psi}\leq \frac{C_{1}}{\varepsilon}$ ($C_{1}$ independent of $\delta$). Then,
\begin{equation}\label{estimateoneovertwo}
\norm{F_{d_{\delta}\geq \delta}e^{-it\widehat{H}(\varepsilon, \delta)}\psi}\leq (2C_{1})^{1/2}\frac{\varepsilon^{1/2}}{\delta}.
\end{equation}
$F_{\cdot}$ indicates the characteristic function of the region indicated.

Moreover, let $\widehat{H}_{D}(\varepsilon, \delta)$ be the Hamiltonian $\widehat{H}(\varepsilon, \delta)$ with Dirichlet boundary conditions on the set $\{(x, y)\in\field{R}{2}: d_{\delta}(x, y) = \delta\}$. 

Let us suppose that $\delta=\delta(\varepsilon)$, $\lim_{\varepsilon\to 0}\delta(\varepsilon)=0$. Taking into account \eqref{estimateoneovertwo}, let us also assume that
\begin{equation}\label{assumptionondelta}
\lim_{\varepsilon\to 0}\frac{\varepsilon^{1/2}}{\delta(\varepsilon)}=0.
\end{equation}

Then, for all $t\in[0, T]$, we have
\begin{equation}\label{estimateoneovereight}
\norm{F_{d_{\delta}<\delta}e^{-it\widehat{H}(\varepsilon, \delta)}\psi - e^{-it\widehat{H}_{D}(\varepsilon, \delta)}F_{d_{\delta}<\delta}\psi}\leq C_{2}(C_{1}, T)\bigg(\frac{\varepsilon^{1/4}}{\delta^{5/2}} + \frac{\varepsilon^{1/2}}{\delta^{3}}\bigg).
\end{equation}
\end{theo}

\begin{rem} The theorem implies that if we choose a $\delta(\varepsilon)$ such that 
\begin{equation}\label{assumptiononeoverten}
\lim_{\varepsilon\to 0}\frac{\varepsilon^{1/10}}{\delta(\varepsilon)}=0
\end{equation}
then we can restrict ourselves to analyze the Dirichlet Hamiltonian $\widehat{H}_{D}(\varepsilon, \delta(\varepsilon))$, which is localized inside the region where tubular coordinates are defined. This, however, means that we have to consider a ``tube'' encircling the graph whose diameter is much bigger than the localization of the transverse states, which for an harmonic oscillator is $\varepsilon^{1/2}$.
\end{rem}

\begin{rem}
As already observed in Froese and Herbst (2001), the estimate \eqref{estimateoneovereight} is not optimal.
\end{rem}

\begin{proof}
Let us first prove \eqref{estimateoneovertwo}. 

Since $\norm{\widehat{H}(\varepsilon, \delta)\psi}\leq \frac{C_{1}}{\varepsilon}$, we have from Schwarz inequality 
\begin{equation*}
<\psi, \widehat{H}(\varepsilon, \delta)\psi>\leq  \frac{C_{1}}{\varepsilon}.
\end{equation*}
This implies immediately that
\begin{equation}\label{energyestimatetwo}
\frac{C_{1}}{\varepsilon}\geq <\widehat{H}^{1/2}\psi, \widehat{H}^{1/2}\psi>= \frac{1}{2}\norm{\nabla\psi}^{2} + \frac{1}{2\varepsilon^{2}}\norm{d_{\delta}\psi}^{2} \Rightarrow \norm{d_{\delta}\psi}^{2} \leq 2C_{1}\varepsilon.
\end{equation}
It follows then 
\begin{equation}
\delta^{2}<F_{d_{\delta}\geq\delta}\psi, F_{d_{\delta}\geq\delta}\psi>\ \leq\ <d_{\delta}F_{d_{\delta}\geq\delta}\psi, d_{\delta}F_{d_{\delta}\geq\delta}\psi>\ \leq\ \norm{d_{\delta}\psi}^{2}\ \leq\ 2C_{1}\varepsilon.
\end{equation}
The same argument can applied also to $e^{-it\widehat{H}(\varepsilon, \delta)}\psi$, so \eqref{estimateoneovertwo} is proved.

We need now to prove an estimate on the behaviour of the gradient of $\psi$ away from the graph. 

Let $\tilde{\chi} \in \textrm{C}^{\infty}_{0}(\field{R}{})$ be $1$ when $1/4<\abs{x}<3/4$ and $0$ when $\abs{x}\leq 1/8$ or $\abs{x}\geq 7/8$, then the function
\begin{equation*}
\chi(u) := \tilde{\chi}\bigg(\frac{1}{2(\alpha-\lambda_{1})}\abs{u} + \frac{1}{4} - \frac{\lambda_{1}}{2(\alpha - \lambda_{1})}\bigg)
\end{equation*}
will be $1$ when $\lambda_{1}<\abs{u}<\alpha$ and $0$ for $\abs{u}$ near zero. If we choose $\lambda_{1}$ and $\alpha$ such that $0<\lambda_{1}<\alpha<\delta$, then $\chi$ is well defined (and $ \in \textrm{C}^{\infty}_{0}(\field{R}{})$) when $u$ is the coordinate along the direction normal to the curve $\Gamma_{\vartheta, \delta}$. 

We have then
\begin{equation*}
\norm{F_{\lambda_{1}<d_{\delta}<\alpha}\nabla\psi}=\norm{F_{\lambda_{1}<d_{\delta}<\alpha}\nabla(\chi\psi)}\leq
\norm{\nabla(\chi\psi)}.
\end{equation*}
Using again the Schwarz inequality and the fact that $\chi\psi\in\mathcal{D}(\Delta)$ (the potential is bounded on the support of $\chi$) we get
\begin{equation*}
\norm{\nabla(\chi\psi)}\leq\norm{\Delta(\chi\psi)}^{1/2}\norm{\chi\psi}^{1/2},
\end{equation*}
so, to estimate $\norm{F_{\lambda_{1}<d_{\delta}<\alpha}\nabla\psi}$ we need to get an estimate on $\norm{\Delta(\chi\psi)}$. To obtain it, we use an energy estimate of second order, i. e., we calculate the quadratic form associated to $\widehat{H}(\varepsilon, \delta)^{2}$.

\begin{equation}\label{hsquare}
\widehat{H}(\varepsilon, \delta)^{2} = \frac{1}{4}\abs{p}^{4} + \bigg(\frac{1}{2\varepsilon^{2}}d_{\delta}^{2}\bigg)^{2} + \sum_{j}p_{j}\frac{1}{2\varepsilon^{2}}d_{\delta}^{2}p_{j} - \frac{1}{2\varepsilon^{2}}\Delta d_{\delta}^{2},
\end{equation}
where $p=-i\nabla$.
The first three terms are positive operators, while if we take the mean value of the last one with respect to the state $\chi\psi$ we get
\begin{equation*}\begin{split}
&<\chi\psi, \Delta d_{\delta}^{2} \chi\psi> = \int_{d_{\delta}<\delta}\, dxdy\, \abs{\chi\psi}^{2}\Delta d_{\delta}^{2} = \int_{\abs{u}<\delta}\, dsdu\, [1+uk_{\delta}(s)]\, \abs{\chi\psi}^{2}\, [1+uk_{\delta}(s)]^{-1}\cdot\\
&\cdot\deriv{u}{}\{[1+uk_{\delta}(s)]2u\} = \int_{\abs{u}<\delta}\, dsdu\, [1+uk_{\delta}(s)]\, \abs{\chi\psi}^{2}\cdot 2 + \int_{\abs{u}<\delta}\, dsdu\, \abs{\chi\psi}^{2}\, 2uk_{\delta}(s)\leq \\
& \leq C\norm{\chi\psi}^{2},
\end{split}
\end{equation*}
where  the Jacobian of the change to curvilinear coordinates is given by 
\begin{equation}
\frac{\partial (q_{1}, q_{2})}{\partial (s, u)} = 1 + k_{\delta}(s)u
\end{equation}
and in the last step we have used the fact that
\begin{equation*}
\sup_{\abs{u}<\delta}\abs{uk_{\delta}(s)}\leq \delta k_{\delta}(s) = \vartheta k\bigg(\frac{s}{\delta}\bigg)\leq const.\ \textrm{(independent of}\ \delta) \quad .
\end{equation*}

Taking the mean value of \eqref{hsquare} with respect to $\chi\psi$ we obtain then
\begin{equation*}
\norm{\frac{1}{2}\Delta(\chi\psi)}^{2}\leq \norm{\widehat{H}(\varepsilon, \delta)(\chi\psi)}^{2} + \frac{C}{\varepsilon^{2}}, 
\end{equation*}
which can be written equivalently as
\begin{equation*}
\frac{1}{2}\norm{\Delta(\chi\psi)}\leq \frac{C^{1/2}}{\varepsilon} + \norm{\widehat{H}(\varepsilon, \delta)\psi} + \frac{1}{2}\norm{[\Delta, \chi]\psi}.
\end{equation*}
The last term is equal to
\begin{equation*}
[\Delta, \chi]\psi = (\Delta\chi)\psi + \nabla\chi\cdot\nabla\psi,
\end{equation*}
and we can estimate its norm changing to curvilinear coordinates,
\begin{equation*}\begin{split}
& \nabla\chi = \deriv{x}{}\tilde{\chi}(x(u))\frac{1}{2(\alpha - \lambda_{1})}\frac{u}{\abs{u}}\boldsymbol{n}(s),\\
& \Delta\chi = (1+k_{\delta}u)^{-1}\deriv{u}{}[(1+k_{\delta}u)\deriv{u}{}\chi] = \deriv{u}{2}\chi + (1+k_{\delta}u)^{-1}k_{\delta}\deriv{u}{}\chi =\\
&= \deriv{x}{2}\tilde{\chi}(x(u))\frac{1}{2(\alpha - \lambda_{1})^{2}} + \deriv{x}{}\tilde{\chi}(x(u))\frac{1}{2(\alpha - \lambda_{1})}\frac{u}{\abs{u}}\frac{k_{\delta}}{1+k_{\delta}u}.
\end{split}
\end{equation*}
Using \eqref{energyestimatetwo} to estimate $\norm{\nabla\psi}$, we have then
\begin{equation}\label{estimatecommutator}
\norm{[\Delta, \chi]\psi}\leq \frac{C}{\varepsilon^{1/2}(\alpha-\lambda_{1})} + \frac{C}{(\alpha-\lambda_{1})^{2}} + \frac{C}{\delta(\alpha-\lambda_{1})}.
\end{equation}
In what follows, we will need to choose $\alpha$ and $\lambda_{1}$ proportional to $\delta$. Assumption \eqref{assumptionondelta} implies then that all terms in \eqref{estimatecommutator} are \emph{at most} of order $\varepsilon^{-1}$.

To sum up, we have
\begin{equation}
\norm{\Delta(\chi\psi)}\leq \frac{C}{\varepsilon},
\end{equation}
from which it follows that (assuming that $\alpha$ and $\lambda_{1}$ are proportional to $\delta$ and that $\delta(\varepsilon)$ satisfies \eqref{assumptionondelta})
\begin{equation}\label{estimatenabla}
\norm{F_{\lambda_{1}<d_{\delta}<\alpha}\nabla\psi}\leq C \varepsilon^{-1/2}\frac{\varepsilon^{1/4}}{\delta^{1/2}} = \frac{C}{\varepsilon^{1/4}\delta^{1/2}}.
\end{equation}

Let now $\tilde{\xi}$ be a function in $\textrm{C}^{\infty}_{0}(\field{R}{})$ such that $\tilde{\xi}(x)=1$ when $\abs{x}<1/4$ and $\tilde{\xi}(x)=0$ when $\abs{x}>1/2$. We define the function $\xi$ by the equation $\xi(u):=\tilde{\xi}(u/\delta)$, where $u$ is the curvilinear coordinate normal to the curve. 

Because of \eqref{estimateoneovertwo}, to prove \eqref{estimateoneovereight} is enough to show that 
\begin{equation*}
\norm{e^{it\widehat{H}_{D}(\varepsilon, \delta)}\xi e^{-it\widehat{H}(\varepsilon, \delta)}\psi - \xi\psi}\leq C_{2}(C_{1}, T)\bigg(\frac{\varepsilon^{1/4}}{\delta^{5/2}}+\frac{\varepsilon^{1/2}}{\delta^{3}}\bigg)
\end{equation*}
for $t\in[0, T]$. Let 
\begin{equation*}
\phi_{t, \varepsilon, \delta} := e^{it\widehat{H}_{D}(\varepsilon, \delta)}\xi e^{-it\widehat{H}(\varepsilon, \delta)}\psi - \xi\psi.
\end{equation*}
Integrating the derivative we have
\begin{equation*}\begin{split}
\phi_{t, \varepsilon, \delta} &= i\int_{0}^{t}\ ds\ e^{is\widehat{H}_{D}(\varepsilon, \delta)}[\widehat{H}_{D}(\varepsilon, \delta)\xi - \xi\widehat{H}(\varepsilon, \delta)]e^{-is\widehat{H}(\varepsilon, \delta)}\psi =\\
&= \int_{0}^{t}\ ds\ e^{is\widehat{H}_{D}(\varepsilon, \delta)}[\nabla\xi\cdot p - (i/2)\Delta\xi]e^{-is\widehat{H}(\varepsilon, \delta)}\psi,
\end{split}
\end{equation*}
therefore
\begin{equation*}
\norm{\phi_{t, \varepsilon, \delta}}^{2} = \int_{0}^{t}\ ds \ <e^{-is\widehat{H}_{D}(\varepsilon, \delta)} \phi_{t, \varepsilon, \delta}, [\nabla\xi\cdot p - (i/2)\Delta\xi]e^{-is\widehat{H}(\varepsilon, \delta)}\psi>.
\end{equation*}
Let now $\tilde{\zeta}$ be a $\textrm{C}_{0}^{\infty}(\field{R}{})$ function which is $1$ on the support of $\deriv{x}{}\tilde{\xi}$ and $0$ when $\abs{x}$ is near zero. As above, we denote by $\zeta(u) := \tilde{\zeta}(u/\delta)$. We can then write
\begin{equation}\label{estimatephi}\begin{split}
&\norm{\phi_{t, \varepsilon, \delta}}^{2} \leq \int_{0}^{t}\ ds\ \norm{\zeta e^{-is\widehat{H}_{D}(\varepsilon, \delta)} \phi_{t, \varepsilon, \delta}}(\norm{\nabla\xi\cdot p e^{-is\widehat{H}(\varepsilon, \delta)}\psi} + \norm{(1/2)\Delta\xi e^{-is\widehat{H}(\varepsilon, \delta)}\psi}) \leq \\
&\leq C\bigg(\frac{1}{\delta^{3/2}\varepsilon^{1/4}} + \frac{1}{\delta^{2}}\bigg)\int_{0}^{t}\ ds\ \norm{\zeta e^{-is\widehat{H}_{D}(\varepsilon, \delta)} \phi_{t, \varepsilon, \delta}},
\end{split}
\end{equation}
where we have used \eqref{estimatenabla} and the definition of $\xi$.

Now
\begin{equation*}\begin{split}
&<\phi_{t, \varepsilon, \delta}, \widehat{H}_{D}(\varepsilon, \delta)\phi_{t, \varepsilon, \delta>}\ \leq\ 2<\xi e^{-it\widehat{H}(\varepsilon, \delta)}\psi, \widehat{H}_{D}(\varepsilon, \delta)\xi e^{-it\widehat{H}(\varepsilon, \delta)}\psi> +\\ 
&+2<\xi\psi, \widehat{H}_{D}(\varepsilon, \delta)\xi\psi> = 2<\xi e^{-it\widehat{H}(\varepsilon, \delta)}\psi, \bigg[-\frac{1}{2}\Delta\xi -i\nabla\xi\cdot p +\xi\widehat{H}_{D}(\varepsilon, \delta)\bigg] e^{-it\widehat{H}(\varepsilon, \delta)}\psi> +\\
&+ 2<\xi \psi, \bigg[-\frac{1}{2}\Delta\xi -i\nabla\xi\cdot p +\xi\widehat{H}_{D}(\varepsilon, \delta)\bigg]\psi>.
\end{split} 
\end{equation*}
Using again equation \eqref{estimatenabla} and the definition of $\xi$, we get
\begin{equation*}\begin{split}
&\abs{<\xi\ \psi, -\frac{1}{2}\Delta\xi\ \psi>} \leq \frac{C}{\delta^{2}},\\
&\abs{<\xi\ \psi, -i\nabla\xi\cdot p\ \psi>} \leq \frac{C}{\varepsilon^{1/4}\delta^{3/2}},\\
&\abs{<\xi\ \psi, \xi\widehat{H}_{D}(\varepsilon, \delta)\ \psi>} \leq \frac{C}{\varepsilon},
\end{split}
\end{equation*}
and corresponding equations with $e^{-it\widehat{H}(\varepsilon, \delta)}\psi$ instead of $\psi$. If we suppose that the sequence $\delta(\varepsilon)$ satisfies \eqref{assumptionondelta}, then all the terms grow at most as $\varepsilon^{-1}$, so we obtain in the end
\begin{equation*}
<\phi_{t, \varepsilon, \delta}, \widehat{H}_{D}(\varepsilon, \delta)\phi_{t, \varepsilon, \delta}>\leq \frac{C}{\varepsilon}.
\end{equation*}
Repeating the proof of \eqref{estimateoneovertwo}, we can then show that
\begin{equation*}
\norm{\zeta e^{-is\widehat{H}_{D}(\varepsilon, \delta)} \phi_{t, \varepsilon, \delta}}\leq \frac{C\varepsilon^{1/2}}{\delta},
\end{equation*}
and substituting this back in \eqref{estimatephi} we get
\begin{equation}
\norm{\phi_{t, \varepsilon, \delta}}^{2}\leq C\bigg(\frac{1}{\delta^{3/2}\varepsilon^{1/4}} + \frac{1}{\delta^{2}}\bigg)\frac{\varepsilon^{1/2}}{\delta}= C\bigg(\frac{\varepsilon^{1/4}}{\delta^{5/2}} + \frac{\varepsilon^{1/2}}{\delta^{3}}\bigg).
\end{equation}
\end{proof}

Now, let us fix a sequence $\delta(\varepsilon)$ satisfying \eqref{assumptiononeoverten}. As in last section, we consider the time evolution of a product state localized inside one of the two edges, away from the vertex,
\begin{equation}\label{psittwoedges}\begin{split}
& \psi_{t}^{\varepsilon} = e^{-it\widehat{H}(\varepsilon, \delta(\varepsilon))}\psi_{0},\\
& \psi_{0}(x, y) = f(x)\Phi_{n}^{\varepsilon}(y),
\end{split}
\end{equation}
where $(x, y)$ is the system of coordinates associated to one of the edges, $f\in\textrm{C}^{\infty}_{0}(\field{R}{})$ and $\Phi_{n}^{\varepsilon}$ has been defined in \eqref{definitionofphin}. If we choose $\varepsilon$ sufficiently small, the tubular coordinates associated to the curve $\Gamma_{\vartheta, \delta}$, $(s_{\delta}, u_{\delta})$, coincide with $(x, y)$ apart from a small neighbourhood of the vertex. The state $\psi_{0}$ is then well defined and independent of $\delta$. The limit $\varepsilon\to 0$ gives us therefore the leading behaviour of an initial state which propagates through a tube which curves slowly with respect to the transverse wavelength. 

Equation \eqref{estimateoneovertwo} allows us to discard $F_{d_{\delta(\varepsilon)}>\delta(\varepsilon)}\psi_{t}^{\varepsilon}$, while \eqref{estimateoneovereight} allows us to approximate $F_{d_{\delta(\varepsilon)}<\delta(\varepsilon)}\psi_{t}^{\varepsilon}$ with $e^{-it\widehat{H}_{D}(\varepsilon, \delta(\varepsilon))}F_{d_{\delta(\varepsilon)}<\delta(\varepsilon)}\psi_{0}$.

We can now prove 
\begin{prop}\label{approximationtwotoone}
Let $\psi_{t}^{\varepsilon}$ be given by \eqref{psittwoedges}, then, for $t\in[0, T]$,
\begin{equation}\begin{split}
&\norm{\exp[-it\widehat{H}_{D}(\varepsilon, \delta(\varepsilon))]F_{d_{\delta(\varepsilon)}<\delta(\varepsilon)}\psi_{0} +\\ &-\exp[-it\widehat{K}(\delta(\varepsilon)) - itE_{n}/\varepsilon](f)\cdot F_{d_{\delta(\varepsilon)}<\delta(\varepsilon)}\Phi_{n}^{\varepsilon}(u_{\delta(\varepsilon)})}\to 0, \quad \varepsilon\to 0,
\end{split}
\end{equation}
where 
\begin{equation}\label{hamkdelta}
\widehat{K}(\delta(\varepsilon)) = -\frac{1}{2}\deriv{s}{2} -\frac{k_{\delta}(s)^{2}}{8}\quad .
\end{equation}
\end{prop}
\begin{proof}
The proof is an application of the fundamental theorem of calculus (also called, in this context, Duhamel formula), 
\begin{equation*}\begin{split}
&\norm{\exp[-it\widehat{H}_{D}(\varepsilon, \delta(\varepsilon))]F_{d_{\delta(\varepsilon)}<\delta(\varepsilon)}\psi_{0} +\\ &-\exp[-it\widehat{K}(\delta(\varepsilon)) - itE_{n}/\varepsilon](f)\cdot F_{d_{\delta(\varepsilon)}<\delta(\varepsilon)}\Phi_{n}^{\varepsilon}(u_{\delta(\varepsilon)})} =\\
&= \norm{\{\exp[it\widehat{H}_{D}(\varepsilon, \delta(\varepsilon))]\exp[-it\widehat{K}(\delta(\varepsilon))-itE_{n}/\varepsilon] - \mathbb{I}\}F_{d_{\delta(\varepsilon)}<\delta(\varepsilon)}f\cdot\Phi_{n}^{\varepsilon}(u_{\delta(\varepsilon)}) +\\ &+F_{d_{\delta(\varepsilon)}<\delta(\varepsilon)}f(x)\Phi_{n}^{\varepsilon}(u_{\delta}) - F_{d_{\delta(\varepsilon)}<\delta(\varepsilon)}f(x)\Phi_{n}^{\varepsilon}(y)}\leq\\
&\leq  \norm{\{\exp[it\widehat{H}_{D}(\varepsilon, \delta(\varepsilon))]\exp[-it\widehat{K}(\delta(\varepsilon))-itE_{n}/\varepsilon] - \mathbb{I}\}F_{d_{\delta(\varepsilon)}<\delta(\varepsilon)}f\cdot\Phi_{n}^{\varepsilon}(u_{\delta})} +\\ &+ \norm{F_{d_{\delta(\varepsilon)}<\delta(\varepsilon)}f(x)\Phi_{n}^{\varepsilon}(u_{\delta}) - F_{d_{\delta(\varepsilon)}<\delta(\varepsilon)}f(x)\Phi_{n}^{\varepsilon}(y)} = \\
&=\norm{\{\exp[it\widehat{H}_{D}(\varepsilon, \delta(\varepsilon))]\exp[-it\widehat{K}(\delta(\varepsilon))-itE_{n}/\varepsilon] - \mathbb{I}\}F_{d_{\delta(\varepsilon)}<\delta(\varepsilon)}f\cdot\Phi_{n}^{\varepsilon}(u_{\delta})},
\end{split}
\end{equation*}
because the second term is zero if $\varepsilon$ (and therefore $\delta$) is sufficiently small.

Applying now Duhamel formula\footnote[1]{For every fixed $\delta$, the domain of $\widehat{K}(\delta)$ is $H^{2}(\field{R}{})$, so $\exp[-it\widehat{K}(\delta)](f)\cdot F_{d_{\delta}<\delta}\Phi(u_{\delta})$ is in the domain of $\widehat{H}_{D}(\varepsilon, \delta)$.}, we have that
\begin{equation}\label{duhameltwo}\begin{split}
&\norm{\{\exp[it\widehat{H}_{D}(\varepsilon, \delta(\varepsilon))]\exp[-it\widehat{K}(\delta(\varepsilon))-itE_{n}/\varepsilon] - \mathbb{I}\}F_{d_{\delta(\varepsilon)}<\delta(\varepsilon)}f\cdot\Phi_{n}^{\varepsilon}(u_{\delta})}\leq \\
&\leq \int_{0}^{t}\ ds\ \norm{[\widehat{H}_{D}(\varepsilon, \delta(\varepsilon)) - \widehat{K}(\delta(\varepsilon)) - E_{n}/\varepsilon]\cdot\\
&\cdot\exp[-is\widehat{K}(\delta(\varepsilon))-isE_{n}/\varepsilon]F_{d_{\delta(\varepsilon)}<\delta(\varepsilon)}f\cdot\Phi_{n}^{\varepsilon}(u_{\delta})}.
\end{split}
\end{equation} 

The Hamiltonian $\widehat{H}_{D}(\varepsilon, \delta(\varepsilon))$ in curvilinear coordinates, acting on $L^{2}(\field{R}{}\times[0, \delta], ds du)$, is given by
\begin{equation*}\begin{split}
\widehat{H}_{D}(\varepsilon, \delta(\varepsilon)) &= -\frac{1}{2}\frac{1}{[1+uk_{\delta}(s)]^{2}}\frac{\partial^{2}}{\partial s^{2}} + \frac{1}{[1+uk_{\delta}(s)]^{3}}uk_{\delta}'(s)\frac{\partial}{\partial s} + V(s, u) +\\
& -\frac{1}{2}\frac{\partial^{2}}{\partial u^{2}} + \frac{1}{2\varepsilon^{2}}u^{2},
\end{split}
\end{equation*}
where $V$ is the geometric potential,
\begin{equation}\label{geometricpotential}
V(s, u)=\frac{1}{2}\bigg\{-\frac{k_{\delta}^{2}}{4[1+uk_{\delta}]^{2}} + \frac{uk_{\delta}''}{2[1+uk_{\delta}]^{2}} - \frac{5}{4}\frac{u^{2}(k_{\delta}')^{2}}{[1+uk_{\delta}]^{2}}\bigg\}.
\end{equation}
Making a unitary dilation by the factor $\varepsilon^{1/2}$ along $u$, we get an operator acting on $L^{2}(\field{R}{}\times [0, \delta(\varepsilon)/\varepsilon^{1/2}], ds du)$, given by
\begin{equation}\label{scaledhamiltonian}\begin{split}
D_{\varepsilon^{1/2}}\widehat{H}_{D}(\varepsilon, \delta(\varepsilon))D_{\varepsilon^{1/2}}^{\dagger} &= -\frac{1}{2}\frac{1}{[1+\varepsilon^{1/2}uk_{\delta}(s)]^{2}}\frac{\partial^{2}}{\partial s^{2}} + \frac{1}{[1+\varepsilon^{1/2}uk_{\delta}(s)]^{3}}\varepsilon^{1/2}uk_{\delta}'(s)\frac{\partial}{\partial s} +\\
&+ V(s, \varepsilon^{1/2}u) -\frac{1}{2\varepsilon}\frac{\partial^{2}}{\partial u^{2}} + \frac{1}{2\varepsilon}u^{2},
\end{split}
\end{equation}
where
\begin{equation*}
D_{\varepsilon^{1/2}}\psi(u) = \varepsilon^{1/4}\psi(\varepsilon^{1/2}u).
\end{equation*}

Therefore, equation \eqref{duhameltwo} becomes 
\begin{equation*}\begin{split}
&\norm{\{\exp[it\widehat{H}_{D}(\varepsilon, \delta(\varepsilon))]\exp[-it\widehat{K}(\delta(\varepsilon))-itE_{n}/\varepsilon] - \mathbb{I}\}F_{d_{\delta(\varepsilon)}<\delta(\varepsilon)}f\cdot\Phi_{n}^{\varepsilon}(u_{\delta})}\leq \\
&\leq \int_{0}^{t}\ ds\ \norm{[D_{\varepsilon^{1/2}}\widehat{H}_{D}(\varepsilon, \delta(\varepsilon))D_{\varepsilon^{1/2}}^{\dagger} - \widehat{K}(\delta(\varepsilon)) - E_{n}/\varepsilon]\cdot\\
&\cdot\exp[-is\widehat{K}(\delta(\varepsilon))-isE_{n}/\varepsilon]f\cdot F_{d_{\delta(\varepsilon)}<\delta(\varepsilon)/\varepsilon^{1/2}}\cdot\Phi_{n}^{\varepsilon=1}(u_{\delta})}\quad .
\end{split}
\end{equation*}

Therefore, it is clear from previous equations that 
\begin{equation*}\begin{split}
& \norm{[V(s, \varepsilon^{1/2}u) + k_{\delta}^{2}/8]\exp[-is\widehat{K}(\delta(\varepsilon))-isE_{n}/\varepsilon]f\cdot F_{d_{\delta(\varepsilon)}<\delta(\varepsilon)/\varepsilon^{1/2}}\cdot\Phi_{n}^{\varepsilon=1}(u_{\delta})} =\\
&= O(\varepsilon^{1/2}/\delta^{3})\to 0, \quad \varepsilon\to 0,
\end{split}
\end{equation*}
\begin{equation*}
\begin{split}
&\bigg\rVert\bigg(-\frac{1}{2\varepsilon}\frac{\partial^{2}}{\partial u^{2}} + \frac{1}{2\varepsilon}u^{2} - \frac{E_{n}}{\varepsilon}\bigg)\exp[-is\widehat{K}(\delta(\varepsilon))-isE_{n}/\varepsilon]f\cdot\\  &\cdot F_{d_{\delta(\varepsilon)}<\delta(\varepsilon)/\varepsilon^{1/2}}\cdot\Phi_{n}^{\varepsilon=1}(u_{\delta})\bigg\lVert \to 0, \quad \varepsilon\to 0,
\end{split}
\end{equation*}
so we need to control only the terms containing the derivative with respect to $s$ in \eqref{scaledhamiltonian}. 

Using lemma \ref{estimatederivatives}, proved below, we have that
\begin{equation*}\begin{split}
&\bigg\lVert\frac{1}{[1+\varepsilon^{1/2}uk_{\delta}(s)]^{3}}\varepsilon^{1/2}uk_{\delta}'(s)\frac{\partial}{\partial s}\exp[-it\widehat{K}(\delta(\varepsilon))-itE_{n}/\varepsilon]f(s)\cdot\\ &\cdot F_{d_{\delta(\varepsilon)}<\delta(\varepsilon)/\varepsilon^{1/2}}\cdot\Phi_{n}^{\varepsilon=1}(u_{\delta})\bigg\rVert\leq (\varepsilon^{1/2}\norm{k_{\delta}'}_{L^{\infty}}\cdot\norm{\deriv{s}{}f}_{L^{2}} + \abs{t}\cdot\varepsilon^{1/2}\norm{k_{\delta}}_{L^{\infty}}\norm{k_{\delta}'}^{2}_{L^{\infty}}\norm{f}_{L^{2}})\cdot\\
&\cdot\norm{u\Phi_{n}^{\varepsilon=1}(u)} = O(\varepsilon^{1/2}/\delta^{5})\to 0, \quad \varepsilon\to 0.
\end{split}
\end{equation*}

In the same way we have also that
\begin{equation*}\begin{split}
&\bigg\lVert-\frac{1}{2}\bigg\{\frac{1}{[1+\varepsilon^{1/2}uk_{\delta}(s)]^{2}}-1\bigg\}\frac{\partial^{2}}{\partial s^{2}}\exp[-it\widehat{K}(\delta(\varepsilon))-itE_{n}/\varepsilon]f(s)\cdot\\ &\cdot F_{d_{\delta(\varepsilon)}<\delta(\varepsilon)/\varepsilon^{1/2}}\cdot\Phi_{n}^{\varepsilon=1}(u_{\delta})\bigg\rVert \leq C\varepsilon^{1/2}\norm{k_{\delta}}_{\infty}[\norm{\deriv{s}{2}f} + \abs{t}\cdot\\ &\cdot (2\norm{k_{\delta}}_{\infty}\cdot\norm{k_{\delta}'}_{\infty}\norm{\deriv{s}{}f} + \norm{k_{\delta}'}_{\infty}^{2}\norm{f} + \norm{k_{\delta}}_{\infty}\norm{k_{\delta}''}_{\infty}\norm{f})] = O(\varepsilon^{1/2}/\delta^{5})\to 0, \quad \varepsilon\to 0.
\end{split}
\end{equation*}
\end{proof}

\begin{rem}
As stressed above, in this model the dynamics under a strong constraining potential is well approximated by Dirichlet boundary conditions on a ``large'' tube surrounding the smooth curve. Proposition \ref{approximationtwotoone} says that this choice gives the same result as the procedure of constraining first the particle to the motion to the curve, and then taking the limit when the curve approaches the graph.
\end{rem}

\begin{lem}\label{estimatederivatives} Let $\widehat{H}$ be the one-dimensional Hamiltonian $\widehat{H} = -\frac{1}{2}\deriv{x}{2} + V$, where $V$ is a potential bounded together with its first two derivatives, then, given $\psi\in H^{1}(\field{R}{})$, we have
\begin{equation}\label{estimatefirstderivative}
\norm{\deriv{x}{}\exp(-it\widehat{H})\psi}_{L^{2}(\field{R}{})}\leq \norm{\deriv{x}{}\psi}_{L^{2}(\field{R}{})} + \abs{t}\cdot \norm{\deriv{x}{}V}_{L^{\infty}}\cdot\norm{\psi}_{L^{2}(\field{R}{})},
\end{equation}  
and, given $\varphi\in H^{2}(\field{R}{})$, 
\begin{equation}\label{estimatesecondderivative}\begin{split}
&\norm{\deriv{x}{2}\exp(-it\widehat{H})\varphi}_{L^{2}(\field{R}{})}\leq \norm{\deriv{x}{2}\varphi}_{L^{2}(\field{R}{})} +\\
&+ \abs{t}(2 \norm{\deriv{x}{}V}_{L^{\infty}}\cdot\norm{\deriv{x}{}\varphi}_{L^{2}(\field{R}{})} + \norm{\deriv{x}{2}V}_{L^{\infty}}\cdot\norm{\varphi}_{L^{2}(\field{R}{})})
\end{split}
\end{equation}
\end{lem}

\begin{proof}
Since $V$ is bounded, the domain of the quadratic form associated to $\widehat{H}$ is $H^{1}(\field{R}{})$, and the time evolution sends it into itself. It makes therefore sense to write, for $\psi\in H^{1}(\field{R}{})$, 
\begin{equation*}\begin{split}
&[-i\deriv{x}{}, e^{-it\widehat{H}}]\psi = e^{-it\widehat{H}}\int_{0}^{t}\ ds\ \deriv{s}{}e^{is\widehat{H}}(-i\deriv{x}{})e^{-is\widehat{H}}\psi =\\ 
&=ie^{-it\widehat{H}}\int_{0}^{t}\ ds\ e^{is\widehat{H}}[\widehat{H}, -i\deriv{x}{}]e^{-is\widehat{H}}\psi = e^{-it\widehat{H}}\int_{0}^{t}\ ds\ e^{is\widehat{H}}\ \deriv{x}{}V\ e^{-is\widehat{H}}\psi,
\end{split}
\end{equation*}
but this implies immediately
\begin{equation*}\begin{split}
& -i\deriv{x}{}e^{-it\widehat{H}}\psi = e^{-it\widehat{H}}(-i\deriv{x}{})\psi + [-i\deriv{x}{}, e^{-it\widehat{H}}]\psi\\
& \Rightarrow \norm{-i\deriv{x}{}e^{-it\widehat{H}}\psi} \leq \norm{-i\deriv{x}{}\psi} + \int_{0}^{t}\ ds\ \norm{\deriv{x}{}V e^{-is\widehat{H}}\psi},
\end{split}
\end{equation*}
which gives \eqref{estimatefirstderivative}. 

Following the same path and noticing that
\begin{equation}
[\widehat{H}, -\deriv{x}{2}] = -\deriv{x}{2}V -2\deriv{x}{}V\cdot\deriv{x}{}
\end{equation}
we get \eqref{estimatesecondderivative}.
\end{proof}

To complete the analysis of this case we need to study the limit of the dynamics $\exp[-it\widehat{K}(\delta(\varepsilon))]$ when $\varepsilon\to 0$. 

The limit of one-dimensional Hamiltonians containing rescaled potentials has been studied in detail in the context of the approximation of singular interactions, like the delta coupling, by short range smooth potentials (Albeverio \emph{et al.} (2005) and references therein). The scaling used by us in \eqref{hamkdelta} however, is not covered in the results presented in Albeverio \emph{et al.}, but it can be analyzed using exactly the same techniques.

The idea is to show convergence in norm of the resolvent of $\widehat{K}(\delta(\varepsilon))$ to the resolvent of the Hamiltonian with Dirichlet boundary conditions in $s=0$. As it is well known (Reed and Simon (1972), theorem VIII.21) this implies strong convergence of the corresponding unitary group. 

One could expect convergence to Dirichlet boundary conditions because the potential $-k_{\delta}^{2}/8$ is a strongly attractive well, which becomes deeper and deeper, but whose range is shorter and shorter. As explained in Englisch and \v Seba (1986) in a different context, we expect this to give rise to Dirichlet boundary conditions. This in particular says that the strong convergence of the unitary group (or the norm resolvent convergence) does not capture the behaviour of the eigenvalues which go to $-\infty$ when $\delta\to 0$, because, even though the ground state of $\widehat{K}(\delta(\varepsilon))$ tends to $-\infty$, its resolvent converges to that of a semibounded operator. This phenomenon has already been illustrated in Gesztesy (1980).

We can now prove
\begin{theo}
Let $[\widehat{K}(\delta(\varepsilon)) - z^{2}]^{-1}$ be the resolvent of $\widehat{K}(\delta(\varepsilon))$, where $\Im z > 0$, then
\begin{equation}
[\widehat{K}(\delta(\varepsilon)) - z^{2}]^{-1}\to [\widehat{K}_{D} - z^{2}]^{-1}, \quad \varepsilon\to 0,
\end{equation}
in the norm of bounded operators on $L^{2}(\field{R}{})$,
where $K_{D}$ is the free Laplacian on $L^{2}(\field{R}{})$ with Dirichlet boundary conditions in $s=0$.
\end{theo}

\begin{proof}
The potential $Q(s) := -k^{2}/8$ is in $L^{1}(\field{R}{})\cap L^{\infty}(\field{R}{})$, so we can apply the dilation technique described in Albeverio \emph{et al.} (1984) (see also Albeverio \emph{et al.} (2005)). Applying lemma $A.1$ of Albeverio \emph{et al.} (1984) we get
\begin{equation}
[\widehat{K}(\delta(\varepsilon)) - z^{2}]^{-1} = G_{z} - A_{\delta}(z)[\delta + B_{\delta}(z)]^{-1}C_{\delta}(z), \qquad \Im z > 0,
\end{equation}
where $G_{z}$ is the free resolvent, with kernel $g_{z}(w)$,
\begin{equation}\begin{split}
&G_{z} := (\widehat{H}_{0} - z^{2})^{-1},\qquad g_{z}(w) := \frac{i}{2z}e^{iz\abs{w}}\\
& \widehat{H}_{0} := -\frac{\partial^{2}}{\partial s^{2}}, \qquad D(\widehat{H}_{0}) = H^{2}(\field{R}{}),
\end{split}
\end{equation}
while $A_{\delta}(z)$, $B_{\delta}(z)$ and $C_{\delta}(z)$ are Hilbert-Schmidt operators with kernels 
\begin{equation}\begin{split}
& A_{\delta}(z, s, r) = g_{z}(s- \delta r)\abs{Q(r)}^{1/2},\\
& B_{\delta}(z, s, r) = -\abs{Q(s)}^{1/2}g_{z}[\delta(s-r)]\abs{Q(r)}^{1/2},\\
& C_{\delta}(z, s, r) = -\abs{Q(s)}^{1/2}g_{z}(\delta s-r).
\end{split} 
\end{equation}

It is not difficult to see (lemma $2.3$ Albeverio \emph{et al.} (1984)) that 
\begin{equation}\begin{split}
& A_{\delta}\to A_{0},\\
& B_{\delta}\to B_{0},\\
& C_{\delta}\to C_{0},
\end{split}
\end{equation}
in Hilbert-Schmidt norm, where $A_{0}$, $B_{0}$ and $C_{0}$ have kernels
\begin{equation}\begin{split}
&A_{0}(z, s, r) = g_{z}(s)\abs{Q(r)}^{1/2},\\
&B_{0}(z, s, r) = -g_{z}(0)\abs{Q(s)}^{1/2}\abs{Q(r)}^{1/2},\\
&C_{0}(z, s, r) = -\abs{Q(s)}^{1/2}g_{z}(-r).
\end{split}
\end{equation}

The operator $B_{0}$ is not invertible on the whole Hilbert space, but it is clear from the expression of the kernel that it actually acts on the one-dimensional subspace, denoted by $\mathcal{H}_{Q}$, generated by the vector $\varphi_{Q}$ given by
\begin{equation}
\varphi_{Q}(s) := \frac{\abs{Q(s)}^{1/2}}{\norm{\ \abs{Q(s)}^{1/2}\ }}_{L^{2}(\field{R}{})}= \frac{\abs{Q(s)}^{1/2}}{\norm{Q(s)}^{1/2}_{L^{1}(\field{R}{})}} \quad .
\end{equation}
So we can write
\begin{equation}
B_{0} = -g_{z}(0)\norm{Q(s)}_{L^{1}(\field{R}{})}\varphi_{Q}<\varphi_{Q}, \cdot>.
\end{equation}
On $\mathcal{H}_{Q}$, $B_{0}$ is invertible and the inverse is given by
\begin{equation}
B_{0}^{-1} = -\frac{1}{g_{z}(0)\norm{Q(s)}_{L^{1}(\field{R}{})}}\varphi_{Q}<\varphi_{Q}, \cdot>.
\end{equation}

Since the operator $C_{0}$ has range equal to $\mathcal{H}_{Q}$ and $A_{0}$ acts non trivially only on $\mathcal{H}_{Q}$, we get that 
\begin{equation*}
[\widehat{K}(\delta(\varepsilon)) - z^{2}]^{-1} \to G_{z} - A_{0}B_{0}^{-1}C_{0},
\end{equation*}
which has a kernel given by
\begin{equation}
g_{z}(s-r) - \frac{g_{z}(s)g_{z}(-r)}{g_{z}(0)}, 
\end{equation}
which is the kernel of the resolvent of the Dirichlet Hamiltonian.
\end{proof}

\footnotesize 
\noindent Albeverio, S., Gesztesy, F., H\o{}egh-Krohn, R. and Holden, H., (with an appendix by P. 

Exner) ``Solvable Models in Quantum Mechanics'', 2nd edition, AMS Chelsea Series 

350 (Providence, R.I., 2005).

\noindent Albeverio, S., Gesztesy, F., H\o{}egh-Krohn, R. and Kirsch, W., ``On point interactions in 

one dimension'', J. Operator Theory, {\bf 12}, 101--126 (1984).

\noindent Amovilli, C., Leys, F. E. and March, N. H., ``Electronic energy spectrum of two-dimensional

 solids and a chain of C atoms from a quantum network model'', J. Math. Chem. {\bf 36}, 

93--112 (2004).

\noindent Belov, V. V., Dobrokhotov, S. Yu. and Tudorovskii, T. Ya., ``Asymptotic solutions of 

nonrelativistic equations of quantum mechanics in curved nanotubes: I. Reduction to 

spatially one-dimensional equations'', Theor. and Math. Phys. {\bf 141}, 1562--1592 

(2004).

\noindent Bornemann, F., ``Homogenization in time of singularly perturbed mechanical systems'', 

Lecture Notes in Mathematics, 1687 (Springer-Verlag, Berlin, 1998).

\noindent Carini, J. P., Londergan, J. T., Mullen, K. and Murdock, D. P., ``Bound states and 

resonances in waveguides and quantum wires'', Phys. Rev. B {\bf 46}, 15538--15541 (1992).

\noindent Carini, J. P., Londergan, J. T., Mullen, K. and Murdock, D. P., ``Multiple bound states 

in sharply bent waveguides'', Phys. Rev. B {\bf 48}, 4503--4515 (1993).

\noindent Duclos, P. and Exner, P., ``Curvature induced bound states in quantum waveguides in 

two and three dimensions'', Rev. Math. Phys. {\bf 7}, 73--102 (1995).

\noindent Englisch, H. and \v Seba, P., ``The Stability of Dirichlet and Neumann Boundary Conditions'', 

Rep. Math. Phys. {\bf 23}, 341--348 (1986).

\noindent Exner, P. and Post, O., ``Convergence of spectra of graph-like thin manifolds'', J. Geom. 

Phys. {\bf 54}, 77--115 (2005).

\noindent Gesztesy, F., ``On the one-dimensional Coulomb Hamiltonian'', J. Phys. A: Math. Gen. 

{\bf 13}, 867--875 (1980).

\noindent Goldstone, J. and Jaffe, R. L., ``Bound states in twisting tubes'', Phys. Rev. B {\bf 45}, 

14100--14107 (1992).

\noindent Kostrykin, V. and Schrader, R., ``Kirchhoff's rule for quantum wires'', J. Phys. A: Math.

 Gen. {\bf 32}, 595--630 (1999).

\noindent Kuchment, P., ``Graph models of wave propagation in thin structures'', Waves in Random 

Media {\bf 12}, R1--R24 (2002).

\noindent Kuchment, P., ``Quantum graphs I. Some basic structures'', Waves in Random Media {\bf 14},

 S107--S128 (2004).

\noindent Kuchment, P., ``Quantum graphs II. Some spectral properties of quantum and 

combinatorial graphs'', J. Phys. A {\bf 38}, 4887--4900 (2005).

\noindent Kuchment, P. and Zeng, H., ``Convergence of Spectra of Mesoscopic Systems Collapsing 

onto a Graph'', J. Math. Anal. Appl. {\bf 258}, 671--700 (2001).

\noindent Kuchment, P. and Zeng, H., ``Asymptotics of Spectra of Neumann Laplacians in Thin 

Domains'', in Advances in Differential Equations and Mathematical Physics, Yu. 

Karpeshina, G. Stolz, R. Weikard, and Y. Zeng (Editors), Contemporary Mathematics 

 387, 199--213, AMS 2003.

\noindent Lang, S., ``Differential and Riemannian Manifolds'' (Springer-Verlag, New York, 1995).

\noindent Londergan, J. T., Carini, J. P. and Murdock, D. P., ``Binding and scattering in two-

dimensional systems'', Lecture Notes in Physics, 60 (Springer-Verlag, Berlin, 1999).

\noindent Melnikov, Yu. B. and Pavlov, B. S., ``Two-body scattering on a graph and applications to 

simple nanoelectronic devices'', J. Math. Phys. {\bf 36}, 2813--2825 (1995).

\noindent Mitchell, K., ``Gauge fields and extrapotentials in constrained quantum systems'', Phys. 

Rev. A {\bf 63}, 042112 (2001).

\noindent Post, O., ``Branched quantum waveguides with Dirichlet boundary conditions: the 

decoupling case'', J. Phys. A: Math. Gen. {\bf 38}, 4917--4931 (2005).

\noindent Post, O., ``Spectral Convergence of Non-Compact Quasi-One-Dimensional Spaces'', arXiv:

 math-ph/0512081, v2 2 Jan 2006.

\noindent Reed, M. and Simon, B., ``Methods of Modern Mathematical Physics. I: Functional 

Analysis'' (Academic Press, New York, 1972).

\noindent Royden, H. L., ``Real Analysis'', 3rd edition (Prentice-Hall, Upper Saddle River, 1988).

\noindent Rubinstein, J. and Schatzman, M., ``Variational Problems on Multiply Connected Thin

Strips I: Basic Estimates and Convergence of the Laplacian Spectrum'', Arch. Rational 

Mech. Anal. {\bf 160}, 271--308 (2001).

\noindent Rudin, W., ``Functional Analysis'' (McGraw-Hill Publishing Co., New York, 1973).

\noindent Ruedenberg, K. and Scherr, C. W., ``Free-Electron Network Model for Conjugated 

Systems. I. Theory'', J. Chem. Phys. {\bf 21}, 1565--1581 (1953).

\noindent Sait\=o, Y., ``The limiting equation for Neumann Laplacians on shrinking domains'', 

Electron. J. Diff. Equations {\bf 2000}, 1--25 (2000).

\noindent Sait\=o, Y., ``Convergence of the Neumann Laplacian on Shrinking Domains'', Analysis {\bf 21}, 

171--204 (2001).

\noindent Teufel, S., ``Adiabatic Perturbation Theory in Quantum Dynamics'', Lecture Notes in 

Mathematics, 1821 (Springer-Verlag, Berlin, 2003).

\end{document}